\documentclass[]{article}
\usepackage{epsfig}
\voffset= - 1.0in
\hoffset= - 1.0in         

\catcode`\@=11
\def\marginnote#1{}

\newcount\hour
\newcount\minute
\newtoks\amorpm
\hour=\time\divide\hour by60
\minute=\time{\multiply\hour by60 \global\advance\minute by-\hour}
\edef\standardtime{{\ifnum\hour<12 \global\amorpm={am}%
        \else\global\amorpm={pm}\advance\hour by-12 \fi
        \ifnum\hour=0 \hour=12 \fi
        \number\hour:\ifnum\minute<10 0\fi\number\minute\the\amorpm}}
\edef\militarytime{\number\hour:\ifnum\minute<10 0\fi\number\minute}

\def\draftlabel#1{{\@bsphack\if@filesw {\let\thepage\relax
   \xdef\@gtempa{\write\@auxout{\string
      \newlabel{#1}{{\@currentlabel}{\thepage}}}}}\@gtempa
   \if@nobreak \ifvmode\nobreak\fi\fi\fi\@esphack}
        \gdef\@eqnlabel{#1}}
\def\@eqnlabel{}
\def\@vacuum{}
\def\draftmarginnote#1{\marginpar{\raggedright\scriptsize\tt#1}}

\def\draft{\oddsidemargin -.5truein
        \def\@oddfoot{\sl preliminary draft \hfil
        \rm\thepage\hfil\sl\today\quad\militarytime}
        \let\@evenfoot\@oddfoot \overfullrule 3pt
        \let\label=\draftlabel
        \let\marginnote=\draftmarginnote
   \def\@eqnnum{(\theequation)\rlap{\kern\marginparsep\tt\@eqnlabel}%
\global\let\@eqnlabel\@vacuum}  }

\def\d{\partial}

\def\bea{\begin{eqnarray}}
\def\eea{\end{eqnarray}}

\def\beq{\begin{equation}}
\def\eeq{\end{equation}}
\def\ba{\beq\new\begin{array}{c}}
\def\ea{\end{array}\eeq}
\def\be{\ba}
\def\ee{\ea}
\def\stackreb#1#2{\mathrel{\mathop{#2}\limits_{#1}}}
\def\Tr{{\rm Tr}}

\makeatletter
\newdimen\normalarrayskip              
\newdimen\minarrayskip                 
\normalarrayskip\baselineskip
\minarrayskip\jot
\newif\ifold             \oldtrue            \def\new{\oldfalse}
\def\arraymode{\ifold\relax\else\displaystyle\fi} 
\def\eqnumphantom{\phantom{(\theequation)}}     
\def\@arrayskip{\ifold\baselineskip\z@\lineskip\z@
     \else
     \baselineskip\minarrayskip\lineskip2\minarrayskip\fi}
\def\@arrayclassz{\ifcase \@lastchclass \@acolampacol \or
\@ampacol \or \or \or \@addamp \or
   \@acolampacol \or \@firstampfalse \@acol \fi
\edef\@preamble{\@preamble
  \ifcase \@chnum
     \hfil$\relax\arraymode\@sharp$\hfil
     \or $\relax\arraymode\@sharp$\hfil
     \or \hfil$\relax\arraymode\@sharp$\fi}}
\def\@array[#1]#2{\setbox\@arstrutbox=\hbox{\vrule
     height\arraystretch \ht\strutbox
     depth\arraystretch \dp\strutbox
     width\z@}\@mkpream{#2}\edef\@preamble{\halign
\noexpand\@halignto
\bgroup \tabskip\z@ \@arstrut \@preamble \tabskip\z@ \cr}%
\let\@startpbox\@@startpbox \let\@endpbox\@@endpbox
  \if #1t\vtop \else \if#1b\vbox \else \vcenter \fi\fi
  \bgroup \let\par\relax
  \let\@sharp##\let\protect\relax
  \@arrayskip\@preamble}
\def\eqnarray{\stepcounter{equation}%
              \let\@currentlabel=\theequation
              \global\@eqnswtrue
              \global\@eqcnt\z@
              \tabskip\@centering
              \let\\=\@eqncr
 \halign to \displaywidth\bgroup
    \eqnumphantom\@eqnsel\hskip\@centering
    $\displaystyle \tabskip\z@ {##}$%
    \global\@eqcnt\@ne \hskip 2\arraycolsep
         $\displaystyle\arraymode{##}$\hfil
    \global\@eqcnt\tw@ \hskip 2\arraycolsep
         $\displaystyle\tabskip\z@{##}$\hfil
         \tabskip\@centering
    &{##}\tabskip\z@\cr}
\begingroup\ifx\undefined\newsymbol \else\def\input#1 {\endgroup}\fi
\newfont{\hr}{msbm10}
\newfont{\ams}{msam10}
\font\teneufm=cmmib10
\font\seveneufm=cmmib7
\font\fiveeufm=cmmib5
\def\bfit#1{{\textfont1=\teneufm\scriptfont1=\seveneufm
\scriptscriptfont1=\fiveeufm
\mathchoice{\hbox{$\displaystyle#1$}}{\hbox{$\textstyle#1$}}
{\hbox{$\scriptstyle#1$}}{\hbox{$\scriptscriptstyle#1$}}}}
\font\numbers=cmss12
\font\upright=cmu10 scaled\magstep1
\def\stroke{\vrule height8pt width0.4pt depth-0.1pt}
\def\topfleck{\vrule height8pt width0.5pt depth-5.9pt}
\def\botfleck{\vrule height2pt width0.5pt depth0.1pt}
\def\Zmath{\vcenter{\hbox{\numbers\rlap{\rlap{Z}\kern 0.8pt\topfleck}\kern
2.2pt
                   \rlap Z\kern 6pt\botfleck\kern 1pt}}}
\def\Qmath{\vcenter{\hbox{\upright\rlap{\rlap{Q}\kern
                   3.8pt\stroke}\phantom{Q}}}}
\def\Nmath{\vcenter{\hbox{\upright\rlap{I}\kern 1.7pt N}}}
\def\Cmath{\vcenter{\hbox{\upright\rlap{\rlap{C}\kern
                   3.8pt\stroke}\phantom{C}}}}
\def\Rmath{\vcenter{\hbox{\upright\rlap{I}\kern 1.7pt R}}}
\def\Z{\ifmmode\Zmath\else$\Zmath$\fi}
\def\Q{\ifmmode\Qmath\else$\Qmath$\fi}
\def\N{\ifmmode\Nmath\else$\Nmath$\fi}
\def\C{\ifmmode\Cmath\else$\Cmath$\fi}
\def\R{\ifmmode\Rmath\else$\Rmath$\fi}

\def\stackreb#1#2{\mathrel{\mathop{#2}\limits_{#1}}}
\def\Tr{{\rm Tr}}

\def\Bf#1{\mbox{\boldmath $#1$}}
\def\balpha{{\Bf\alpha}}

\def\bmu{{\Bf\mu}}
\def\bphi{{\Bf\phi}}

\def\bsigma{{\bfit\sigma}}

\def\d{\partial}

\def\Im{{\rm Im}}
\def\Re{{\rm Re}}

\def\2{{1\over 2}}
\def\N2{${\cal N}=2$}
\def\4N{${\cal N}=4$}
\def\1N{${\cal N}=1$}

\def\beq{\begin{equation}}
\def\eeq{\end{equation}}
\def\ba{\beq\new\begin{array}{c}}
\def\ea{\end{array}\eeq}
\def\be{\ba}
\def\ee{\ea}
\def\stackreb#1#2{\mathrel{\mathop{#2}\limits_{#1}}}

\begin{document}


\begin{flushright}
FIAN/TD-03/02\\
ITEP/TH-08/02\\
PNPI/2467/02\\
hepth/0202172
\end{flushright}

\vfil

\begin{center}
\baselineskip20pt
{\bf \LARGE Non-Abelian Confinement via Abelian Flux Tubes\\
in Softly Broken \N2 SUSY QCD}
\end{center}
\bigskip
\begin{center}
\baselineskip12pt
{\large A.~Marshakov}\\
\medskip
{\em Theory Department, Lebedev Physics Institute, and\\
Institute of Theoretical and Experimental Physics, Moscow, Russia}\\
{\sf e-mail:\ mars@lpi.ru, andrei@heron.itep.ru}\\
\bigskip
{\large A.~Yung}\\
\medskip
{\em Petersburg Nuclear Physics Institute, Gatchina, St.
Petersburg, and\\
Institute of Theoretical and Experimental Physics, Moscow, Russia}\\
{\sf e-mail:\ yung@thd.pnpi.spb.ru}\\
\end{center}
\bigskip\bigskip\medskip

\begin{center}
{\large\bf Abstract} \vspace*{.2cm}
\end{center}

\begin{quotation}
We study  confinement in softly broken \N2 SUSY QCD
with gauge group $SU(N_c)$ and $N_f$ hypermultiplets of fundamental matter
(quarks) when the Coulomb branch is lifted by small mass of adjoint matter.
Concentrating mostly on the theory with $SU(3)$ gauge group
we discuss the \1N vacua which arise in the weak coupling at large
values of quark masses and study flux tubes and monopole confinement in
these vacua. In particular we find the BPS strings in $SU(3)$ gauge
theory formed by two interacting $U(1)$ gauge
fields and two scalar fields generalizing ordinary Abrikosov-Nielsen-Olesen
vortices. Then we focus on the $SU(3)$ gauge theories with
$N_f=4$ and $N_f=5$
flavors with equal masses. In these theories there are \1N vacua
with restored $SU(2)$ gauge subgroup in quantum theory since
$SU(2)$ subsectors are not asymptotically free. We show that
although the confinement in these theories is due to Abelian flux
tubes the multiplicity of meson spectrum is the same as expected
in a theory with non-Abelian confinement.
\end{quotation}

\vfil

\newpage

\setcounter{footnote}{0}
\setcounter{equation}{0}

\section{Introduction}

According to Mandelstam, Polyakov and 't Hooft \cite{MPH} confinement of
charges arises as Meissner effect upon condensation of
monopoles. Once monopoles  condense the electric flux is confined
within the electric flux tube \cite{A,NO}
connecting the heavy trial charge and anti-charge.
The flux tube has
constant energy per unit length -- the string tension
$T=(2\pi\alpha')^{-1}$,
and this ensures that confining potential between heavy charge and
anti-charge increases linearly with their separation. However,
since dynamics of monopoles is hard to control in
non-supersymmetric gauge theories,
this picture of confinement for many years remained to be an
unjustified qualitative scheme.

The breakthrough in this direction was made by Seiberg and
Witten in \cite{SW1,SW2}. Constructing exact solution to \N2
supersymmetric gauge theory they have shown that the condensation of
monopoles really occurs near the monopole point in the
moduli space of the theory once \N2 supersymmetry is broken
down to \1N by the mass term of adjoint matter \cite{SW1}.
After the work of Seiberg and Witten it has become very
important to understand to what extent this Abelian confinement of electric
charges  is similar to confinement of color we expect
(but cannot control) in real QCD.
Moreover, one expects the QCD-like confinement also in \1N
supersymmetric QCD which can be obtained as a limit of large
mass of the adjoint matter $\mu$ of softly broken \N2 QCD, but again, we have
no control on the dynamics of this theory in the large $\mu$ limit.

One important distinction noticed by Douglas and Shenker \cite{DS} appears
in $SU(N_c)$ gauge theories with $N_c \ge 3$. Since $SU(N_c)$ gauge group
is broken down to $U(1)^{N_c-1}$ by the VEV's of adjoint scalars
there are $N_c-1$ generically different
flux tubes, one per each $U(1)$ factor. Numerous flux tubes  lead to
existence of too many hadronic states in the spectrum~\cite{DS} (see also
\cite{HSZ} for the D-brane reinterpretation of this result). In particular,
there are $N_c$  different  sets of quark-antiquark meson
Regge trajectories \footnote{Each set corresponds
to main trajectory together with "daughter" trajectories.}.
For example, in pure $SU(N_c)$ gauge
theory the number of
families of sets of these trajectories with different slope
is the integer part of
$(N_c+1)/2$. The presence of many quark-antiquark meson trajectories
reflects the essentially Abelian nature of confinement in Seiberg-Witten theory.

On the other hand in \1N supersymmetric theories described
in the framework of Seiberg's duality \cite{S}
we should get non-Abelian confinement
since without adjoint fields breaking of the gauge symmetry down to
Abelian subgroup does not occur. It is believed that condensation
of "magnetic quarks" of non-Abelian dual theory leads to confinement
of ordinary quarks. Still the mechanism of this confinement is
not understood. The problem is that usually non-Abelian dual
gauge groups, say $SU(r)$ groups do not admit flux tubes.

The bridge between two approaches was suggested in \cite{APS,konishi}.
It was shown that  certain \1N vacua of softly broken \N2 QCD
at small $\mu$ preserve non-Abelian $SU(r)$ subgroups of $SU(N)$
gauge symmetry  ($r<N_c$). The theories at
some of these vacua correspond at large $\mu$ to \1N QCD
described by Seiberg's duality. This suggests that confinement
at these vacua should be non-Abelian even at small $\mu$.
In particular we should have only one set of
quark-antiquark meson Regge trajectories.
In this paper we are going to study the outlined above proposal for
the non-Abelian
confinement from the side of softly broken \N2 QCD at small $\mu$.
In particular we analyse what happens to Abelian flux tubes
when a non-Abelian subgroup of gauge symmetry is restored at
certain \1N vacua.
We will mostly focus on theory with the $SU(3)$ gauge group
and $N_f \le 5$ flavors of fundamental hypermultiplets to be called as
quarks below.
In sect.~\ref{ss:vacua} we review the vacuum structure of the theory
studied partially in
\cite{APS,konishi}. We consider the case of large
quark masses $m_A$, $A=1,\ldots N_f$, to keep theory at weak coupling
and discuss vacua where quarks develop VEV's. These vacua are
classified by the
number ${\tt r}$ of different colors for which quarks have non-zero
VEV's. For $SU(3)$ gauge theory the isolated vacua can have ${\tt r}=0,1,2$
and we are interested mostly in  weak coupling
vacua with ${\tt r}=1$ and/or ${\tt r}=2$ for large values of $m_A$.

In sect.~\ref{ss:spectrum} we present weak coupling Abelian description
of the low energy effective theory around these vacua and discuss
the low energy spectra.
Generically the $SU(3)$ gauge group is broken down to its maximal
$U(1)\times U(1)$ Abelian subgroup
by the VEV's of adjoint matter and we have two light photon
multiplets.
In sect.~\ref{ss:tubes} and sect.~\ref{ss:strings} we study in detail
 magnetic
flux tubes in ${\tt r}=1$ and ${\tt r}=2$ vacua
responsible for the confinement of monopoles. It turns out that
${\tt r}=1$ vacua
possess the standard Abrikosov-Nielsen-Olesen (ANO) flux tubes \cite{A,NO}
described by one $U(1)$ gauge field and one scalar field.
In the limit of small masses of the adjoint matter $\mu$
these flux tubes are BPS-saturated \cite{B}.
However, at ${\tt r}=2$ vacua the flux tubes turn out to have more complicated
structure. They are formed now by two gauge fields interacting
with two scalar fields. We consider the lattice of these flux tubes
of more general type and study which ones among them are BPS
states. To do this we  calculate the interaction potential of
strings at large distances.

In sect.~\ref{ss:nonabelian} we consider finally the ${\tt r}=2$ vacua in
the $SU(3)$ gauge theories with $N_f=4$ and $N_f=5$ in special limit
of equal quark masses, when the $SU(2)$ subgroup
of gauge symmetry is restored. The reason is that classically
restored $SU(2)$ subgroup stays unbroken also at quantum level
since corresponding $SU(2)$ subsectors
are not asymptotically free (though, of course, the $SU(3)$ gauge theory
with $N_f=4$ and $N_f=5$ is asymptotically free itself). We study what happens
to the Abelian flux tubes and monopole confinement
under restoration of the $SU(2)$ gauge subgroup
and find that one of flux tubes becomes unstable
and eventually disappears from the spectrum. Two other
"elementary" strings become essentially identical in this limit.
This eliminates unwanted
multiplicity of the hadron spectrum mentioned above since in this
vacuum we get only one set of meson Regge trajectories as expected in
a theory with non-Abelian confinement.

\section{Vacuum structure
\label{ss:vacua}}

\subsection{Superpotential and vacuum equations}

Consider softly broken \N2 $SU(N_c)$ gauge theory with $N_f$ flavors.
The field content
in terms of \1N supermultiplets can be described by the gauge multiplet
$W_\alpha$ and chiral multiplet $\Phi$ in the adjoint representation
(forming together \N2 gauge supermultiplet) so that first contains the
gauge field $\left(A_{\mu}\right)^{i}_{j}$ and complex Weyl fermion
$\left(\lambda_{\alpha}\right)^{i}_{j}$,
while the second -- complex scalar $\Phi^{i}_{j}$ and complex Weyl fermion
$\left(\psi_{\alpha }\right)^{i}_{j}$, ($i,j=1,\dots,N_c$), -- all being
$N_c\times N_c$ matrices with zero trace and 2$N_f$ chiral multiplets
of matter $Q^{A,i}$ and ${\tilde Q}_{A,i}$ in the $N_c$ and ${\bar N}_c$
representations respectively, i.e. $i=1,\dots,N_c$ and $A=1,\dots,N_f$.
The vacuum structure we are going to discuss can be associated with
extrema $\delta{\cal W}$ of the superpotential
\be
\label{supot}
{\cal W} =
\sum_{A=1}^{N_f}\left(\sqrt{2}\tilde Q_{A,i}\Phi^{i}_{j}Q^{A,j} + m_A\tilde
Q_{A,i}Q^{A,i}\right) + \sum_{k=2}^{N_c-1}\mu_k\Tr\Phi^k
\ee
Here the last term generally breaks \N2 supersymmery
down to ${\cal N}=1$ and we have chosen it to be linear in
the independent invariant combinations $\Tr\,\Phi^k$. In fact, later on
we simplify it further assuming that only the coefficient $\mu_2\equiv\mu$
in front of the mass term of adjoint matter is non-zero.

To find vacua of the theory we also have to impose
the D-term condition which reduces to  $[\Phi^{\dagger},\Phi]=0$ and
\be
\label{Dterm}
D^{i}_{j} \equiv \sum_{A=1}^{N_f}\left(Q^{A,i}{\bar Q}_{A,j} -
{\bar{\tilde Q}}^{A,i}{\tilde Q}_{A,j}\right) = \nu\delta^{i}_{j}
\ee
The parameter $ \nu = \nu (Q,\tilde Q,\bar Q,{\bar{\tilde Q}})
=\frac1{N_c}\Tr D$ is in fact
some
function of quark's
VEV's and it can be nonzero only if {\em all} $Q_j\neq 0$ for
$j=1,\dots,N_c$.
The variation of the superpotential (\ref{supot}) gives rise to the
following set of equations
\be
\label{eqs}
\sqrt{2}\Phi^{i}_{j}Q^{A,j} + m_AQ^{A,i} = 0\ \ \ \ \forall\ A=1,\dots,N_f
\\
\sqrt{2}\tilde Q_{A,i}\Phi^{i}_{j} + m_A \tilde Q_{A,j} = 0
\ \ \ \ \forall\ A=1,\dots,N_f
\ee
including also $F$-term condition ${\d{\cal W}\over\d\Phi_{ij}}=0$,
which should be taken into account together with (\ref{Dterm}). One should
also remember that $\Tr\Phi=0$, say, introducing the Lagrange multiplier
$\mu_1$ into (\ref{supot}), the same is to impose vanishing condition
only onto the traceless part
\be
\label{vanFterm}
\sqrt{2}\sum_{A=1}^{N_f}\left(Q^{A,i}\tilde Q_{A,j}-{\delta^{i}_{j}\over N_c}
\left(\sum_k Q^{Ak}\tilde Q_{Ak}\right)\right) +
\sum_{k\geq 2}k\mu_k\left( \left(\Phi^{k-1}\right)^{i}_{j}
-{\delta^{i}_{j}\over N_c}\Tr\Phi^{k-1}\right) = 0
\ee
of $ \sqrt{2}F^{i}_{j} +
\sum_{k\geq 2}k\mu_k\left(\Phi^{k-1}\right)^{i}_{j}$
with
\be
\label{Fterm}
F^{i}_{j}\equiv \sum_{A=1}^{N_f}Q^{A,i}\tilde Q_{A,j}.
\ee
Let us first investigate the structure of solutions to vacuum equations
(\ref{Dterm}), (\ref{eqs}) and (\ref{vanFterm}). We shall basically use
quasiclassical regime
when all masses of the quarks $m_A \gg \Lambda_{QCD}$ where the coupling is
small and quasiclassics is a good approximation. Whenever it is
not possible we will add extra arguments based on the
Seiberg-Witten analysis of strong coupling regime of SUSY gauge
theories \cite{SW1,SW2}.

Equation  $[\Phi^{\dagger},\Phi]=0$ gives rise immediately to the conclusion
that, in general position, up to gauge transformations
\be
\label{diag}
\Phi = \left(
\begin{array}{cccc}
  \phi_1 & 0 & \dots & 0 \\
  0 & \phi_2 & \dots & 0 \\
  \vdots &  & \ddots & \vdots \\
  0 &  & \dots & \phi_{N_c}
\end{array}\right)
\\
\Tr\Phi = \sum_{i=1}^{N_c}\phi_i = 0
\ee
and gauge group $SU(N_c)$ is broken down to $U(1)^{N_c-1}$ since the
off-diagonal vector fields $\left(A_{\mu}\right)^{i}_{j}$ acquire the masses proportional to
\be
\label{Higgs}
[\Phi, A_\mu]^{i}_{j} =(\phi_i-\phi_j)\left(A_{\mu}\right)^{i}_{j}
\ee
due to the Higgs effect. When $\phi_i=\phi_j$ for some $i\neq j$ non-Abelian
gauge symmetry may be partially restored. For diagonal $\Phi$ (\ref{diag})
the second term in (\ref{vanFterm}) vanishes for $i\neq j$
and one ends up
with a simple problem of counting of the eigenvectors of matrix $\Phi$. We
shall consider in detail the case of gauge group $SU(3)$.

\subsection{SU(3) gauge group
\label{ss:su3}}

The $SU(3)$ \1N SUSY gauge theory without matter has exactly $N_c=3$ vacua,
all in the strong coupling regime. This is a particular case of general
situation for an $SU(N_c)$ pure gauge theory, which has $N_c$ points in the
moduli space when $N_c-1$ monopoles become massless (see, for example,
\cite{KLTY,AF,DS}).

Let us add one flavor with mass $m_A=m$
so that the equations (\ref{eqs}) and linear combinations of the
equations (\ref{vanFterm}) turn into
\be
\label{eqssu31}
(\sqrt{2}\phi_1 + m)Q^1=0 \ \ \ \ \ \ {\tilde Q}_1(\sqrt{2}\phi_1 + m)=0
\\
(\sqrt{2}\phi_2 + m)Q^2=0 \ \ \ \ \ {\tilde Q}_2(\sqrt{2}\phi_2 + m)=0
\\
(-\sqrt{2}\phi_1-\sqrt{2}\phi_2 + m)Q^3=0 \ \ \ \ \
{\tilde Q}_3(-\sqrt{2}\phi_1-\sqrt{2}\phi_2 + m)=0
\\
\sqrt{2}\left({\tilde Q}_1Q^1 - {\tilde Q}_3Q^3\right) =
- (2\mu_2-3\mu_3\phi_2) (2\phi_1+\phi_2)
\\
\sqrt{2}\left({\tilde Q}_2Q^2 - {\tilde Q}_3Q^3\right) =
- (2\mu_2-3\mu_3\phi_1) (\phi_1+2\phi_2)
\ee
where we have restricted ourselves to (the only essential in $SU(3)$ case)
nonzero $\mu_2$ and $\mu_3$.
Matrix $\Phi$ may have
only one eigenvector, which can be by gauge transformation turned
to any given direction in colour space, for example
\footnote{In the case of the $SU(3)$ gauge group we will use common notations
$u$, $d$ and $s$ for the $Q^i$ with $i=1,2,3$.}
\be
\label{q1}
Q^{A,i} \equiv Q^i = u\delta^{i}_{1} = u\left(
\begin{array}{c}
  1 \\
  0 \\
  0
\end{array}\right)
\ee
and
\be
\label{tq1}
{\tilde Q}_{A,i} \equiv {\tilde Q}_i = {\tilde u}\delta^{1}_{i} =
{\tilde u} \left(
\begin{array}{ccc}
  1 & 0 & 0
\end{array}\right)
\ee
 Taking into account (\ref{q1}), (\ref{tq1}) equations
(\ref{eqssu31}) give rise to $\sqrt{2}\phi_1 + m = 0$
or $\phi_1=-{m\over\sqrt{2}}$.
Then, from the last equation of (\ref{eqssu31}) and vanishing of $Q_2$ and
$Q_3$ one gets that $\phi_2 = - {\phi_1\over 2} = {m\over 2\sqrt{2}}$,
and, as a
consequence $\phi_3 = -\phi_1-\phi_2 = {m\over 2\sqrt{2}}$. As a result
matrix $\Phi$ acquires the form
\be
\label{phir1}
\Phi = {1\over\sqrt{2}}\left(
\begin{array}{ccc}
  -m & 0 & 0 \\
  0 & {m\over 2} & 0 \\
  0 & 0 & {m\over 2}
\end{array}\right)
\ee
It means that the gauge group is in fact broken only up to $SU(2)\times
U(1)$. The vector fields $A^{12}_\mu$ and $A^{13}_\mu$ (and their
superpartners) become very heavy, of the mass $\sim m \gg \Lambda$, while
the mass of the photon $A^{11}_\mu$ and its superpartners is totally defined
by "higgsing" via $Q_1$ and, from the fourth equation of (\ref{eqssu31}), is
given by
\be
\label{mlight}
m^{\rm light} \sim g\sqrt{m\left(\mu_2-
\frac{3}{4\sqrt{2}}\mu_3
m\right)}\
\ee
i.e. is of the scale of SUSY
breaking for generic $\mu_2$ and $\mu_3$
\footnote{In this paper we do not consider
possible complications related to the "fine-tuning"
of (\ref{mlight}), say when for $\mu_3 = {4\sqrt{2}\mu_2\over
3m}$ extra fields could become massless, at least classically.}.
In (\ref{mlight}) we
have also introduced the coupling constant $g$ which comes from
kinetic term and corresponds to correct normalization of mass as
a pole in the propagator. From the equations (\ref{eqssu31}) and
formulas (\ref{q1}) and (\ref{tq1}) we immediately find the value
of the quark condensate $\langle{\tilde Q}Q\rangle$ in this
vacuum
\be
\label{r1q}
\langle{\tilde u}u\rangle = {3\over 2}\
 m \left(\mu_2-\frac{3}{4\sqrt{2}}\mu_3 m\right)
\equiv \frac{\zeta}{2},
\\
\langle{\tilde d}d\rangle = \langle{\tilde s}s\rangle = 0,
\ee
where we introduced parameter $\zeta=2\langle{\tilde u}u\rangle $
The $SU(2)$
subgroup is unbroken at the scale of order of $m$, it interacts only with very
heavy matter $Q_2$ and $Q_3$ (of $m\gg\Lambda$) and by the Seiberg-Witten
mechanism it runs to the strong-coupled phase and has two Seiberg-Witten
vacua \cite{SW1,SW2}. It means that vacua of the full theory would
correspond to
\be
\label{phisu2}
\Phi = {1\over\sqrt{2}}\left(
\begin{array}{ccc}
  -m & 0 & 0 \\
  0 & {m\over 2}\pm \Lambda_{SU(2)} & 0 \\
  0 & 0 & {m\over 2} \mp \Lambda_{SU(2)}
\end{array}\right)
\ee
where $\Lambda_{SU(2)}^4 m = \Lambda_{SU(3)}^5\equiv\Lambda^5$.
The gauge fields $A^{23}_\mu$ and their superpartners would get
masses of the order of $\Lambda_{SU(2)} \ll m$ and the unbroken
$U(1)$ subgroup of this $SU(2)$ will remain light, with the mass
$\sim\sqrt{\mu_2\Lambda_{SU(2)}}$. This
mass is even less than (\ref{mlight}). In order to
avoid running of the $SU(2)$ subgroup into the strong coupling
one has to add more flavors to our gauge theory.

\subsection{More flavors with SU(3)
\label{ss:moresu3}}

Let us now turn to the situation with more flavors.
Of course we can still have vacua of the type considered in previous
section for each flavor, however, in this case more complicated
vacua when different flavors get simultaneously non-zero VEV's also
arise.

From (\ref{eqs}) it follows that either $Q^{A}$ and
$\tilde{Q}_A$ are zero
for a given $A$ or they are  eigenvectors of the diagonal matrix
(\ref{diag}) with eigenvalue $- m_{A}/\sqrt{2}$.
 An important restriction on extra vacua
also comes from the formula (\ref{vanFterm}) with $i\neq
j$ (see (\ref{diag}))
\be
\label{orth} F^{i}_{j} =
\sum_{A=1}^{N_f}Q^{A,i}\tilde Q_{A,j} = 0, \ \ \ \ \ \ i\neq j
\ee
Combining these conditions together we see that
for example, in the case $N_f=2$ one may consider
only $Q^{A,i}=Q^i\delta^{Ai}$, i.e.
\be
\label{q2}
Q^{1,i} = u^1\delta^{i1} = u^1\left(
\begin{array}{c}
  1 \\
  0 \\
  0
\end{array}\right)
\\
Q^{2,i} = d^2\delta^{i2} = d^2\left(
\begin{array}{c}
  0 \\
  1 \\
  0
\end{array}\right)
\ee
and ${\tilde Q}_{A,i}={\tilde Q}_i\delta_{Ai}$, or
\be
\label{tq2}
{\tilde Q}_{1,i} = {\tilde u}_1\delta_{i1} =
{\tilde u}_1 \left(
\begin{array}{ccc}
  1 & 0 & 0
\end{array}\right)
\\
{\tilde Q}_{2,i} = {\tilde d}_2\delta_{i2} =
{\tilde d}_2 \left(
\begin{array}{ccc}
  0 & 1 & 0
\end{array}\right)
\ee
and equations (\ref{eqs}) give rise to
$\sqrt{2}\phi_1 + m_1 = 0$ and $\sqrt{2}\phi_2 + m_2 = 0$
implying
\be
\label{r2phi}
\Phi = -{1\over\sqrt{2}}\left(
\begin{array}{ccc}
  m_1 & 0 & 0 \\
  0 & m_2 & 0 \\
  0 & 0 & -m_1-m_2
\end{array}\right)
\ee
This is a general situation up to
gauge rotations in the colour space, so
one gets an extra vacuum with two
$U(1)$ gauge groups softly broken at "light" level of (\ref{mlight}).
Indeed, from (\ref{q2}), (\ref{tq2}) and, following from (\ref{vanFterm})
relations one finds the following values for the vacuum condensates:
\be
\label{vacr2}
\langle{\tilde u}_1u^1\rangle = -{1\over\sqrt{2}}(2\mu_2-3\mu_3\phi_2)
(2\phi_1+\phi_2) =
(\mu_2+\frac{3}{2\sqrt{2}}\mu_3m_2) (2m_1+m_2)\
\\
\langle{\tilde d}_2d^2\rangle = -{1\over\sqrt{2}}(2\mu_2-3\mu_3\phi_1)
(\phi_1+2\phi_2)
= (\mu_2+ \frac{3}{2\sqrt{2}}\mu_3m_1) (m_1+2m_2)
\ee
The spectrum of light fields will be discussed in detail in
sect.~\ref{ss:r2spec}.
The total number of vacua is in agreement with the formula from \cite{konishi}
\be
\label{konishi}
\#({\rm vacua}) = \sum_{{\tt r}=0}^{\min (N_c-1,N_f)} (N_c-r)C^{N_f}_{\tt r}
\ee
where ${\tt r}$ counts the number of nontrivial eigenvectors or solutions to
(\ref{eqs}). Indeed, ${\tt r}=0$ term gives $N_c$ ($=3$ for $SU(3)$)
Seiberg-Witten vacua "without matter", ${\tt r}=1$ term adds
$(N_c-1)\cdot N_f$
($=2N_f$ for $SU(3)$) vacua, corresponding to (\ref{phisu2}) for each
flavor) -- with the gauge group broken to $SU(N_c-1)$ at the scale
$m\gg\Lambda$ in weak coupling. The term with ${\tt r}=2$
\be
\label{nrvac}
(N_c-2)C^{N_f}_2\ \stackreb{N_c=3}{=}\
{N_f(N_f-1)\over 2}
\ee
corresponds
exactly to the situation, considered in this section.

Formula (\ref{konishi}) has a simple physical meaning.
The factor $(N_c-{\tt r})$ is exactly the
Witten index of unbroken by the
adjoint matter gauge group (in the case we
consider adjoint matter always breaks $SU(N_c)$ down to some
"lower" $SU(N_c-{\tt r})$).  The combinatorial factor $C_{\tt r}^{N_f}$
counts  the number of possibilities to arrange quark VEV's
within $N_f$ flavors.

\subsection{Colliding vacua and Higgs branches
\label{ss:colide}}

If some of the masses of the matter multiplets coincide ($m_A=m_B$ for
$A\neq B$; $A,B=1,\dots,N_f$) one gets a Higgs branch with VEV's $\langle
Q\rangle\neq 0$, "growing" from the corresponding point on the moduli space
of the Coulomb branch. The (real) dimension of the Higgs branch is $4{\cal
H}$, i.e. the dimension of the hyper K\"ahler manifold is four times the
 "number of
hypermultiplets" ${\cal H}$.

In the ${\tt r}=1$ case one can always choose the nontrivial eigenvectors of
$\Phi$ along some fixed direction in the colour space, say only $Q^{1A}\neq
0$ and $\tilde Q_{A1}\neq 0$ for any $A$. Then we get $4N_f$ real
parameters (for coinciding all $m_A=m$ and matrix $\Phi$ of the form of
(\ref{phir1})), which should obey two real F-term and
one real D-term relations.
One extra degree is "eaten up" by the $U(1)$ gauge group so that
finally the dimension of the Higgs branch appears to be $4N_f-3-1=4(N_f-1)$
or ${\cal H}_{{\tt r}=1}=N_f-1$.

In the ${\tt r}=2$ case the situation is a bit more complicated.
The solutions for
nontrivial $Q$ and $\tilde Q$ can be now chosen in the form of the following
rectangle matrices
\be
\| Q^{kA} \| = \left(
\begin{array}{cccc}
  Q^{11} & Q^{12} & \dots & Q^{1N_f} \\
  Q^{21} & Q^{22} & \dots & Q^{2N_f} \\
  0 & 0 & \dots & 0
\end{array}\right)
\ee
and
\be
\|\tilde Q_{Ak} \| = \left(
\begin{array}{ccc}
  \tilde Q_{11} & \tilde Q_{12} & 0 \\
  \tilde Q_{21} & \tilde Q_{22} & 0 \\
  \vdots & \vdots & \vdots \\
  \tilde Q_{N_f1} & \tilde Q_{N_f2} & 0
\end{array}\right)
\ee
consisting of the eigenvectors of $\Phi = {\rm diag}(-m,-m,2m)$, with $8N_f$
parameters. These parameters obey now four real D-term conditions (\ref{Dterm})
and eight real F-term conditions and gauge group $SU(2)\times U(1)$ "eats"
four "phases", so that the total number of independent parameters is
$8N_f-12-4=8(N_f-2)$ or ${\cal H}_{{\tt r}=2}=2(N_f-2)$, which coincides
with the formula of \cite{APS} for ${\tt r}=2$.

\subsection{Baryonic branches}

Formula (\ref{konishi}) is based on counting of the number $r$ of nonzero
eigenvectors -- solutions to (\ref{eqs}). We have considered the cases
${\tt r}=1$ and ${\tt r}=2$ for the $SU(3)$ gauge theory, but it is obvious
that the maximal number of eigenvectors is ${\tt r}=N_c=3$. This possibility
appears first time for
$N_f=N_c$ ($=3$ if we still consider the $SU(3)$ gauge group) and
is only possible, however, in the situation when $\sum_{A=1}^{N_f}m_A =
\sum_{A=1}^{3}m_A =0$.

Consider, for example, ${\tt r}=2$ vacuum of the previous section.
The relation (\ref{orth}) still requires $Q^{A,i}=Q^i\delta^{Ai}$ and
${\tilde Q}_{A,i}={\tilde Q}_i\delta_{Ai}$, but now for $i,A=1,\dots,3$.
In this way we get the set of vacua, parameterized by
\be
\label{bary}
\sum_{A=1,2}\left({\tilde Q}_{A,1}Q^{A,1} -
{\tilde Q}_{A,3}Q^{A,3}\right)=
{\tilde u}_1u^1 - {\tilde s}_3s^3 =
(\mu_2+\frac{3}{2\sqrt{2}}\mu_3m_2) (2m_1+m_2)
\\
\sum_{A=1,2}\left({\tilde Q}_{A,2}Q^{A,2} -
{\tilde Q}_{A,3}Q^{A,3}\right)=
{\tilde d}_2d^2 - {\tilde s}_3s^3 =
(\mu_2+\frac{3}{2\sqrt{2}}\mu_3m_1) (m_1+2m_2)
\ee
The system (\ref{bary}) describes the baryonic branch
\footnote{To "symmetrize" (\ref{bary}) one may also add a relation
\be
\sum_{A=1,2}\left({\tilde Q}_{A,1}Q^{A,1} -
{\tilde Q}_{A,2}Q^{A,2}\right)=
{\tilde u}_1u^1 - {\tilde d}_2d^2 =
(\mu_2+\frac{3}{2\sqrt{2}}\mu_3m_3) (m_1-m_2)
\ee
which is not independent, but is just the
difference of two in (\ref{bary}).},
indeed (\ref{vacr2}) is
simply obtained from (\ref{bary}) putting $Q_3={\tilde Q}_3=0$, and in this
point $m_3$ can be "unfrozen" from $-(m_1+m_2)$. This is a baryonic branch
since the VEV's of baryons
\be
B = u^1d^2s^3 = {1\over 3!}\epsilon_{ijk}\epsilon_{ABC} Q^{A,i}Q^{B,j}Q^{C,k}
\\
{\tilde B} = {\tilde u}_1{\tilde d}_2{\tilde s}_3 = {1\over 3!}
\epsilon^{ijk}\epsilon^{ABC} {\tilde Q}_{A,i}{\tilde Q}_{B,j}{\tilde Q}_{C,k}
\ee
are nonzero.
The dimension of this baryonic branch is
$\#(Q_i,\tilde Q_i)-\#(F)-\#(D)-\#({\rm phases}) = 12-4-2-2=4$.

The simplest analog of this branch exists already for the $SU(2)$ gauge
theory with $N_f=N_c=2$ matter hypermultiplets with $m_1=-m_2=m$. The
corresponding moduli space is described by
\be
\label{phisu2m}
\Phi = {1\over\sqrt{2}}\left(
\begin{array}{cc}
  -m & 0 \\
  0 & m
\end{array}\right)
\ee
and
\be
Q^{kA} = \left(
\begin{array}{cc}
  Q^1 & 0 \\
  0 & Q^2
\end{array}\right)
\\
{\tilde Q}_{Ak} = \left(
\begin{array}{cc}
  \tilde Q_1 & 0 \\
  0 & \tilde Q_2
\end{array}\right)
\ee
or eight real parameters and the relations (\ref{Dterm}) and (\ref{vanFterm})
 give
one and two real relations on them correspondingly (the last one
\be
Q^1\tilde Q_1 - Q^2\tilde Q_2 = 2\mu m
\ee
is
a particular case of counted carefully general situation when the $i=j$
relations of
(\ref{vanFterm}) impose $2\cdot{\rm rank} = 2\cdot(N_c-1)$ conditions) plus a
$U(1)$ phase, so that the dimension is $8-2-1-1=4$. The corresponding baryon
operators are $B=Q^1Q^2$ and $\tilde B=\tilde Q_1\tilde Q_2$.

One may also add more flavors, up to $N_f=4$ (the conformal point for the
$SU(2)$ theory) when the baryonic branch can be described in terms of
(\ref{phisu2m}) and
\be
Q^{kA} = \left(
\begin{array}{cccc}
  Q^1 & 0 & Q^3 & 0\\
  0 & Q^2 & 0 & Q^4
\end{array}\right)
\\
{\tilde Q}_{Ak} = \left(
\begin{array}{cc}
  \tilde Q_1 & 0 \\
  0 & \tilde Q_2 \\
  \tilde Q_3 & 0 \\
  0 & \tilde Q_4
\end{array}\right)
\ee
modulo (\ref{Dterm}) and (\ref{vanFterm}) which gives $16-2-1-1=12$ parameters
or three hypermultiplets.

\section{Low energy mass spectrum
\label{ss:spectrum}}

In this section we work out the action of the effective
 low energy Abelian theory and use it to study the low energy  spectrum
in  isolated \1N charge vacua of the theory
at $m_{A}\gg \Lambda$ (i.e. at different values
of quark masses when there are no Higgs branches). We put
 $\mu_k=\mu\delta_{k2}$ or keep only $\mu_2\equiv\mu\neq 0$,
assuming also that $\mu\ll \Lambda$.

\subsection{Abelian description}

Let us  consider scales of order $\sqrt{\mu m_{A}}$ which are
well below W-boson masses at small $\mu$. At this scale $SU(3)$ gauge
group is broken down to $U(1)^2$ by VEV of the adjoint scalar (\ref{diag})
\be
\label{phigen}
\Phi =
\left(
\begin{array}{ccc}
  \phi_1 & 0 & 0 \\
  0 & \phi_2 & 0 \\
  0 & 0 & \phi_3
\end{array}\right) =
\frac{1}{2}\left(
\begin{array}{ccc}
  a_3+\frac{a_8}{\sqrt{3}} & 0 & 0 \\
  0 & -a_3+\frac{a_8}{\sqrt{3}} & 0 \\
  0 & 0 & -2\frac{a_8}{\sqrt{3}}
\end{array}\right) \equiv \lambda_3a_3 + \lambda_8a_8
\ee
at generic values at quark masses. We also have two $U(1)$ gauge fields
(in the orthogonal basis denoted
$A^{(3)}_{\mu}$ and $A^{(8)}_{\mu}$; note that the orthogonal basis is
normalized so that $\Tr\Phi^2 = \2\left(a_3^2 + a_8^2\right)$)
introduced via
\be
\label{Amu}
A_{\mu} =
\frac{1}{2}\left(
\begin{array}{ccc}
  A^{(3)}_{\mu}+\frac{A^{(8)}_{\mu}}{\sqrt{3}} & 0 & 0 \\
  0 &   -A^{(3)}_{\mu}+\frac{A^{(8)}_{\mu}}{\sqrt{3}} & 0 \\
  0 & 0 & -2\frac{A^{(8)}_{\mu}}{\sqrt{3}}
\end{array}\right) = \lambda_3 A^{(3)}_{\mu}+
\lambda_8 A^{(8)}_{\mu}
\ee
where our notations correspond to expanding  gauge and adjoint fields
\begin{figure}[tb]
\epsfysize=9cm
\centerline{\epsfbox{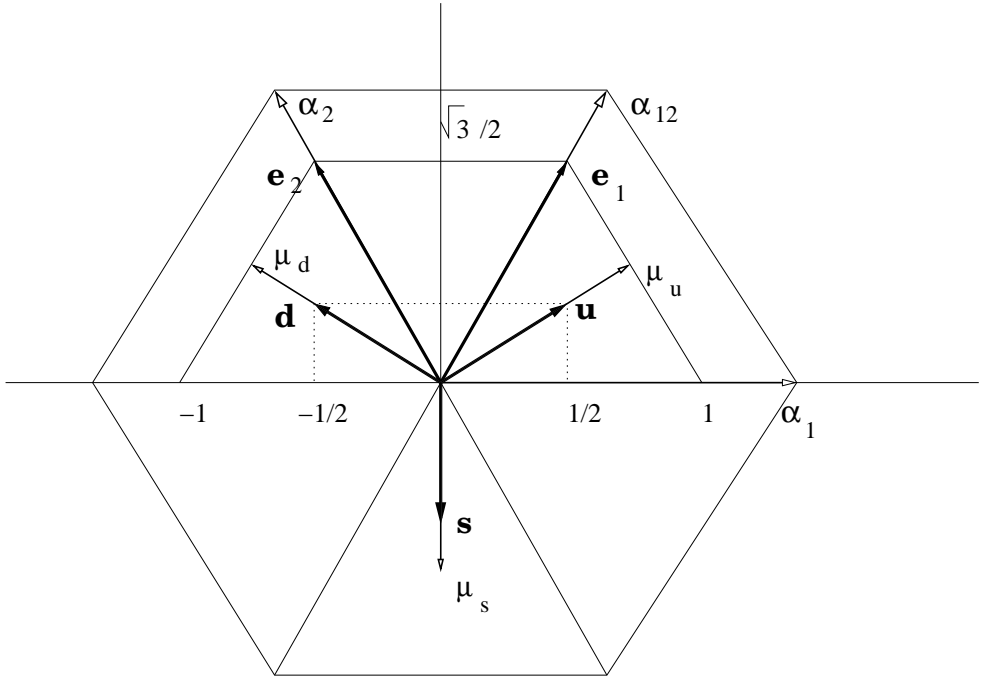}}
\caption{\sl Root and weight vectors in the Cartan plane for
$SU(3)$ group.  We have depicted explicitly the root vectors
$\balpha_1$ and $\balpha_2$ (the simple roots), the highest root
$\balpha_{12}=\balpha_1+\balpha_2$ and the weights of the
fundamental representation $\bmu_1\equiv\bmu_u$,
$\bmu_2\equiv\bmu_d$ and $\bmu_3\equiv\bmu_s$, corresponding to
$u$, $d$ and $s$ quarks respectively. We have also depicted the
"normalized" roots (of unit length) ${\bf e}_1 =
\balpha_{12}/\sqrt{2}$ and ${\bf e}_2 = \balpha_{2}/\sqrt{2}$ as
well as normalized weights ${\bf u}$, ${\bf d}$ and ${\bf s}$
of the length $1/\sqrt{3}$.}
\label{fi:su3}
\end{figure}
in the orthogonal  basis of the Gell-Mann matrices -- the basis of diagonal
Cartan generators of the $SU(3)$ Lie algebra (see (\ref{lam3}) and
(\ref{lam8}) in Appendix). It will become clear later that it is
more adequate to use
 the basis given by (divided by two)  root vectors in the
Cartan subalgebra of $SU(3)$ $\balpha_{12}/2$ and $\balpha_{2}/2$
\footnote{We will also used "normalized" roots
${\bf e}_0 = \balpha_1/\sqrt{2}$
${\bf e}_2 = \balpha_2/\sqrt{2}$ and
${\bf e}_1=\balpha_{12}/\sqrt{2}$ and "normalized" weights, e.g.
${\bf u} = \bmu_1/\sqrt{2}$.},
see Appendix and fig.~\ref{fi:su3},
but we find it technically simpler to work in the
orthogonal basis. The price to pay for this is non-integer electric
charges of $u$ and $d$ quarks with respect to orthogonal
gauge fields, say $A_{\mu}^{(3)}$ and $A_{\mu}^{(8)}$, see below.

In these notations the bosonic part of the low energy effective
Abelian theory acquires the form
\be
\label{qed}
S_{QED}=\int d^4x \left(\frac1{4g^2}\left(F^{(3)}_{\mu\nu}\right)^2 +
\frac1{4g^2}\left(F^{(8)}_{\mu\nu}\right)^2
+
\frac1{g^2}\left|\partial_{\mu}a_3\right|^2 +\frac1{g^2}
\left|\partial_{\mu}a_8\right|^2+ \right.
\\
 + \sum_{A=1}^{N_f}\sum_{i=u,d,s}\left(\left|\nabla^{(i)}_\mu Q_i^{A}\right|^2 +
\left.\left|\nabla^{(i)}_\mu {\tilde Q}_i^{A}\right|^2\right)
+V(u,d,s,a_3,a_8)\right)
\ee
Here $u,d,s$ label different {\em colors} of scalar components
of quark supermultiplets $Q^i$ while
\be
\label{nabla}
\nabla^{(u,d)}_\mu=\partial_\mu-{i\over\sqrt{3}}\; A^{(u,d)}_{\mu} \\
\nabla^{(s)}_\mu=\partial_\mu+{i\over\sqrt{3}}\; A^{(u)}_{\mu}
+{i\over\sqrt{3}}\; A^{(d)}_{\mu}
\ee
where we introduced also (non-orthogonal) components of the gauge fields
interacting directly with $u$- and $d$-quarks
\be
\label{Aud}
A^{(u)}_{\mu}=\frac{\sqrt{3}}{2}A^{(3)}_{\mu}+\frac12 A^{(8)}_{\mu}
\\
A^{(d)}_{\mu}=-\frac{\sqrt{3}}{2}A^{(3)}_{\mu}+\frac12 A^{(8)}_{\mu}
\ee
The potential in the Lagrangian (\ref{qed})
\be
\label{pot0}
V(u,d,s,{\bf a}) = \\
= g^2\sum_{\balpha\in\Delta_+}\left({1\over 4}|D_\balpha|^2 +
\left|{\d{\cal W}\over\d\Phi_\balpha }\right|^2\right) + g^2{\bf D}^2
+ g^2\left|{\d{\cal W}\over\d{\bf a}}\right|^2 +
\sum_{A=1}^{N_f}\sum_{i=u,d,s}\left(\left|{\d{\cal W}\over\d Q^{A,i}}
\right|^2 +
\left|{\d{\cal W}\over\d \tilde Q_{A,i}}\right|^2\right) = \\
= \frac{g^2}{4}\sum_{i\ne j}|D^i_{j}|^2 +
{g^2\over 12}\sum_{i<k}\left(D^i_{i}-D^k_{k}\right)^2 + \frac{g^2}{2}
\sum_{i\ne j}{\d{\cal W}\over\d \Phi^i_{j}}
\overline{\left({\d{\cal W}\over\d \Phi^j_{i}}\right)} + \\
+ g^2\left(\left|{\d{\cal W}\over\d a_3}\right|^2 +
\left|{\d{\cal W}\over\d a_8}\right|^2\right) +
\sum_{A=1}^{N_f}\sum_{i=u,d,s}\left(\left|{\d{\cal W}\over\d Q^{A,i}}
\right|^2 +
\left|{\d{\cal W}\over\d \tilde Q_{A,i}}\right|^2\right)
\ee
is given by D and F terms
(\ref{Dterm}), (\ref{Fterm})) (see also formulas (\ref{eqs}),
(\ref{vanFterm})), expressed in terms of the scalar components
$Q^1\equiv u$, $Q^2\equiv d$, $Q^3\equiv s$ and $a_{3,8}$ (\ref{phigen})
\be
V(u,d,s,a_3,a_8)= \frac{g^2}{4}\sum_{i\ne j}D^i_{j} D^j_{i} +
 g^2\sum_{i\ne j}\bar{F}^i_{j} F^j_{i} +
\\
+ \frac{g^2}{8}\left(D^u_{u}- D^d_{d}\right)^2 + \frac{g^2}{24}
\left(D^u_{u}+ D^d_{d} -2D^s_{s}\right)^2+
\\
+ \frac{g^2}{2}\left|F^u_{u}- F^d_{d} +\sqrt{2}\mu a_3\right|^2+
\frac{g^2}{2}\left|\frac{1}{\sqrt{3}}\left(F^u_{u}+ F^d_{d} -2F^s_{s}\right)+
\sqrt{2}\mu a_8\right|^2 +
\\
\sum_{A=1}^{N_f}\left(\frac12\left|a_3+\frac{a_8}{\sqrt{3}}
+\sqrt{2}m_A \right|^2\left(|u^A|^2 + |\tilde{u}_A|^2\right)+
\frac12\left|-a_3+\frac{a_8}{\sqrt{3}}
+\sqrt{2}m_A \right|^2\left(|d^A|^2 + |\tilde{d}_A|^2\right)+\right.
\\
\left. +  \frac12\left|\frac{2}{\sqrt{3}}a_8+\sqrt{2}m_A \right|^2
\left(|s^A|^2 + |\tilde{s}_A|^2\right)\right)
\label{pot}
\ee
where color indices $i,j$ as in (\ref{qed}) run over the set of $u$, $d$
and $s$.

The QED coupling constant $g$ in (\ref{qed}) is small at small $\mu$,
determined by the mass of light quarks and photons (cf. eq. (\ref{mlight}))
\be
\label{qedg}
{1\over g^2}\sim \log {\mu m\over\Lambda^2}
\ee
With logarithmic accuracy we do not distinguish between two different
coupling constants associated with two $U(1)$ factors in (\ref{qed}).
Below we consider different types of \1N vacua and analyse what kind of
flux tubes  do we have in each type.

\subsection{r=1 vacua
\label{ss:r1vac}}

Consider first ${\tt r}=1$ vacuum when only one $Q^{A,i}=Q^{1,i}$ quark flavor
develops
VEV. We drop flavor index below in this subsection because all other
${\tt r}=1$ vacua have the similar structure. As we have learnt in
sect.~\ref{ss:su3} the VEV's of scalar fields in this vacuum are
given by (\ref{r1q})
where we assume for simplicity that $\mu $ and $m$ are real and positive
and choose the phase of the $u$-quark condensate to vanish.
The  adjoint scalar develop VEV given by (\ref{phir1}), in terms of
fields $a_3$ and $a_8$ this reads
\be
\label{r1a}
\langle a_3\rangle =- \frac{3}{2\sqrt{2}}m \\
\langle a_8\rangle = - \frac{\sqrt{3}}{2\sqrt{2}}m
\ee
From the kinetic term for $u$-quark in (\ref{qed}) one can get the mass
matrix for the $U(1)$ gauge fields $(A^{(3)}_{\mu},A^{(8)}_{\mu})$
\be
\label{r1phmat}
{\cal M}_{\gamma}^2 = \frac{3}{2}g^2\mu\, m\left(
\begin{array}{cc}
  1 & \frac1{\sqrt{3}}  \\
  \frac1{\sqrt{3}} & \frac13  \\
  \end{array}\right)
\ee
whose eigenvalues are
\be
\label{r1mass}
m_1^2=2g^2\mu\ m =\frac23 g^2\zeta
\ee
together with
\be
\label{r1ml}
m_2^2 = 0
\ee
where we have used parameter $\zeta=2\langle\tilde{u}u\rangle = 3\mu m$.
These equations show that one massive photon
appears since $u$-quark develops nonzero VEV (\ref{r1q}) and breaks one of
two $U(1)$ gauge groups while the massless photon is associated
with the second unbroken $U(1)$ group. This fact we have already
pointed out in sect.~\ref{ss:su3}. Classically  the $SU(2)$
group which includes the  latter $U(1)$ factor remains unbroken, see for
example formula (\ref{phir1}). However, in quantum theory this
$SU(2)$ subgroup is broken down due to the Seiberg-Witten mechanism
\cite{SW1,SW2} and the second photon acquires small mass of the order
of $\sqrt{\mu\Lambda_{SU(2)}}$ due to monopole/dyon condensation
in this $SU(2)$ sector.

Now let us consider the mass matrix for the scalars $a_3$ and $a_8$, to the
leading order in $\mu/m$ it can be read off the two last lines in potential
(\ref{pot}). In this approximation it coincides with the photon mass
matrix (\ref{r1phmat}) and therefore one complex $a$-field
remains massless while the other one acquires the same mass
(\ref{r1mass}) as one of the photons.
This is directly related to the fact that \N2 supersymmetry
is {\em not} broken in the
effective low energy QED (\ref{qed}) in the leading order in $\mu/m$
\cite{HSZ,VY}. To see this note, that in this approximation the
perturbation of superpotential (\ref{supot}) proportional to $\mu$
is linear in fluctuations of $a$-fields. Thus it boils down to
the Fayet-Iliopoulos (FI) $F$-term which does not break
\N2 supersymmetry in the effective QED \cite{HSZ,VY}. In the next to
leading order in $\mu/m$  (which corresponds to taking into account
fluctuations of the $a$-fields in the third line of potential (\ref{pot}))
\N2 supersymmetry is broken and \N2 supermultiplets split \cite{VY}.
Below we restrict ourselves to the leading order in $\mu/m$ so that \N2
supersymmetry is preserved in the effective low energy Abelian theory
(\ref{qed}).

To complete the study of the mass spectrum in vicinity of ${\tt r}=1$ vacuum
consider the mass matrix for quarks. The $d$- and $s$-quarks are very heavy
(with masses of the order of $m\gg\Lambda$) and, hence, decouple from
the low-energy theory (see last two terms in
(\ref{pot})). The mass matrix for remaining four real components
of $u$-quark $(\Re u, \Re \tilde{u}, \Im u, \Im \tilde{u})$ can be read
off the terms
$$
\frac{g^2}{6}(D^u_{u})^2
+ \frac{g^2}{2}\left|F^u_{u} +\sqrt{2}\mu a_3\right|^2+
\frac{g^2}{2}\left|\frac{F^u_{u}}{\sqrt{3}}+
\sqrt{2}\mu a_8\right|^2 = {2g^2\over 3}
\left(\left|F^u_{u} -{3\over 2}\mu m\right|^2 + {1\over 4}(D^u_{u})^2\right)
$$
of the potential (\ref{pot}) and has the form
\be
{\cal M}_{u}^2 = 2g^2\mu\, m\left(
\begin{array}{cccc}
  1 & 0 & 0 & 0  \\
  0 & 1 & 0 & 0 \\
  0 & 0 & \frac12 & \frac12 \\
  0 & 0 & \frac12 & \frac12
  \end{array}\right)
\ee
with one zero eigenvalue corresponding to the state "eaten"
by Higgs mechanism, and three other eigenvalues coinciding with
the photon mass (\ref{r1mass}). These three real states of $u$-quark
combine with two real states of massive $a$-field and three states
of massive photon to form the bosonic part of one {\em long} \N2
supermultiplet which contains eight boson and eight fermion states.
The long multiplet appears because
electric charges are screened in the broken $U(1)$ sector by
Higgs condensate, therefore the corresponding central charges of \N2 algebra
vanish, and short BPS multiplets cannot appear \cite{VY}.
To sum up, one gets one long \N2 multiplet with mass (\ref{r1mass})
formed by the $u$-quark condensation and another
one with much smaller mass of the order of
$\sim \sqrt{\mu\Lambda_{SU(2)}}$ formed
by the monopole/dyon  condensation in the classically unbroken $SU(2)$ sector.

\subsection{r=2 vacua in the low energy effective theory
\label{ss:r2spec}}

Let us consider now the ${\tt r}=2$ vacua of sect.~\ref{ss:moresu3}. Assume for
simplicity that we have only two quark flavors with masses $m_1$ and
$m_2$. In this vacuum  quark fields develop VEV's (\ref{vacr2}) while
adjoint VEV's are given by (\ref{r2phi}). In $a_3$, $a_8$ basis they are
\be
\label{r2vev}
\langle a_3\rangle = - \frac{1}{\sqrt{2}}(m_1-m_2) \\
\langle a_8\rangle =-\sqrt{\frac32}(m_1+m_2)
\ee
The mass matrix for the gauge fields $(A^{(3)}_{\mu},A^{(8)}_{\mu})$ can be
read off the kinetic terms for $u$- and $d$-quarks in (\ref{qed}) and
has the form
\be
\label{r2phmat}
{\cal M}_{\gamma}^2 = \frac{g^2\xi}{2}\left(
\begin{array}{cc}
  1+\omega & \frac1{\sqrt{3}}(1-\omega)  \\
  \frac1{\sqrt{3}}(1-\omega) & \frac13(1+\omega)  \\
  \end{array}\right)
\ee
where we have introduced the parameters of ${\tt r}=2$ vacua
\be
\label{xi}
\xi \equiv 2\langle \tilde{u}_1u^1\rangle = 2\mu (2m_1+m_2)
\ee
and
\be
\label{x}
\omega \equiv \frac{\langle\tilde{d}_2d^2\rangle}
{\langle\tilde{u}_1u^1\rangle}=\frac{2m_2+m_1}{2m_1+m_2}
\ee
analogous to the parameter $\zeta$ of ${\tt r}=1$ vacuum (\ref{r1q}).
Two eigenvalues of this mass matrix are given by
\be
\label{r2mass}
\left(m_{\gamma}^2\right)_\pm = \frac{g^2\xi}{2}\Omega_\pm
\ee
with
\be
\label{lam}
\Omega_\pm = \frac23\left( 1+\omega\pm \sqrt{1-\omega+\omega^2}\right)
\ee
We see that for generic values of $m_1$ and $m_2$ both $U(1)$ groups
are broken and both photons acquire masses.
The mass matrix for  two complex fields $a_3$ and $a_8$ is identical
to (\ref{r2phmat}) in the leading order in $\mu/m$ as
can be seen again from (\ref{pot}).

The mass matrix for quarks is now of the size $8\times 8$ including four
(real) components of $u^1$-quark and four components of $d^2$-quark.
It has two zero eigenvalues
associated with the two states ``eaten'' by the Higgs mechanism for two
$U(1)$ gauge factors and two non-zero eigenvalues coinciding with photon
masses (\ref{r2mass}). Each of these non-zero eigenvalues corresponds
to three quark eigenvectors.
Altogether we have two long \N2 multiplets with masses (\ref{r2mass}),
each one containing eight bosonic and eight fermionic states.

Now let us briefly comment on the special case of coinciding masses
$m_1=m_2$ ($\omega=1$) to be considered in detail in
sect.~\ref{ss:nonabelian}. In this case $\langle a_3 \rangle = 0$
and $SU(2)$ subgroup of the $SU(3)$ gauge group is
restored on the Coulomb branch at zero $\mu$ at least classically,
see (\ref{r2phi}). However when one switches on the terms proportional to
$\mu$ this $SU(2)$ subgroup becomes broken completely by $u$- and $d$-quark
condensates. It is easy to see that
all three masses of $SU(2)$ gauge fields are the same and
given by (\ref{r2mass}) with $\Omega_+ = 2$ which corresponds to the special
value $\omega=1$ in (\ref{lam}) when mass matrix becomes diagonal.

\section{Flux tubes
\label{ss:tubes}}

In this section we consider flux tubes in isolated \1N charge vacua of
the theory at $m_{A}\gg \Lambda$ and $\mu\ll \Lambda$.
In this regime the low energy effective theory
reduces to Abelian model (\ref{qed}) at weak coupling so one can
use the semi-classical methods to study string solutions.

\subsection{ANO strings in r=1 vacua
\label{ss:anor1}}

At ${\tt r}=1$ vacuum $d$- and $s$-quarks are heavy and we can simply ignore them.
The VEV's of light fields are given by eqs. (\ref{r1q}) and (\ref{r1a}),
in particular $u$- and $\tilde{u}$-quarks have the same VEV's.
This suggest that one can look for the ANO string solution
using the ansatz (\ref{r1q}), (\ref{r1a}), i.e. to fix
\be
\label{ans01}
d=s=\tilde{d}=\tilde{s}=0
\\
a_3=-\frac{3}{2\sqrt{2}}m, \; a_8=-\frac{\sqrt{3}}{2\sqrt{2}}m
\ee
and express $u$ and $\tilde{u}$
in terms of a single complex field $\varphi$
\be
\label{ansu}
u=\bar{\tilde{u}}= \frac{\varphi}{\sqrt{2}}
\ee
Since $u$-quark interacts only with particular combination $A^{(u)}_{\mu}$
of the gauge fields $A^{(3)}_{\mu}$ and $A^{(8)}_{\mu}$ given by
(\ref{Aud}) it is natural to assume that only this combination
is non-zero on corresponding string solution. To implement this let us
rotate the fields $A^{(3)}_{\mu}$ and $A^{(8)}_{\mu}$ to another
orthogonal basis
\be
\label{r1bas}
A^{(u)}_{\mu}=\frac{\sqrt{3}}{2}A^{(3)}_{\mu}+\frac12 A^{(8)}_{\mu}
\\
A_{2\mu}=-\frac12 A^{(3)}_{\mu} +\frac{\sqrt{3}}{2}A^{(8)}_{\mu}
\ee
The meaning of the subscript ``2'' of $A_{2\mu}$ means that it is directed
along $\balpha_2\sim{\bf e}_2$ (see fig.~\ref{fi:su3}) and
at the moment we only need that
\be
\label{zeroA}
A_{2\mu}=0
\ee
on our string solution. It is easy to check that the ansatz
(\ref{ansu}), (\ref{zeroA}) satisfies the equations of motion.
With this ansatz the bosonic part of softly broken \N2 QED (\ref{qed})
reduces to the form of standard Abelian Higgs model (the relativistic
version of the Landau-Ginzburg model)
\be
\label{ah}
S_{\rm AH}=\int {\rm d}^4x\left(\frac1{4g^2}\left(F^{(u)}_{\mu\nu}\right)^2
 +|\nabla^{(u)}_\mu \varphi|^2+\lambda\left(
|\varphi|^2-\zeta\right)^2\right)\,
\ee
with particular
value of quartic coupling and electric charge $n_e=1/\sqrt{3}$ of
the field $\varphi$, see (\ref{nabla}).
The field $\varphi$ develops VEV (\ref{r1q})
$\left|\langle\varphi\rangle\right|^2=\zeta=
3|\mu m|$, therefore U(1) gauge group is broken, or the
photon acquires mass $m^2_{\gamma}=2g^2\zeta/3$ while the
Higgs mass is equal to  $m_H^2=4\lambda\zeta$.
Note,  that the photon mass
coincides with that of (\ref{r1mass}), since the fields
$A^{(u)}_{\mu}$ and $A_{2\mu}$ diagonalize the photon mass matrix
(\ref{r1phmat}), if it were not so the fields $A^{(u)}_{\mu}$ and $A_{2\mu}$
would mix leading to contradiction between condition (\ref{zeroA})
and equations of motion.

Strictly speaking the substitution $\langle a_3\rangle =
-\frac{3}{2\sqrt{2}}m$ and
$ \langle a_8\rangle =- \frac{\sqrt{3}}{2\sqrt{2}}m$ does not
satisfy equations of motion following from (\ref{qed}).
In fact VEV's $a_3$ and $a_8$ (\ref{ans01}) get
$x$-dependent corrections of the order of $\sqrt{\mu/m}$ \cite{HSZ,VY}.
However, as we already explained in sect.~\ref{ss:r1vac}
one can neglect this effect in the leading order in $\mu/m$ and
consider $\zeta$ to be just a constant $\zeta=3\mu m$. In this approximation
the perturbation term in the superpotential (\ref{supot})
is linear in $a$ and reduces to the FI $F$-term which does not break
\N2 supersymmetry in effective QED \cite{HSZ,VY}.
As is explained
in detail in \cite{VY} the ANO strings in $U(1)$ theory
are BPS-saturated in this limit (see also
\cite{FG}). Below we use the term "BPS string" for these
"almost BPS" solutions, they belong to the short \N2 multiplets which become
long \1N multiplets when next to leading order
corrections (breaking \N2 supersymmetry) are taken into
account \cite{VY}.

Let us now briefly remind the basic features of the BPS ANO strings
\cite{B}.
For arbitrary $\lambda$ in (\ref{ah}) the Higgs mass $m_H$ (the
inverse correlation
length) and photon mass $m_\gamma$ (the
inverse penetration depth) are different
and their ratio is important parameter in theory of superconductivity,
characterizing the type of
superconductor. Namely, for $m_H<m_\gamma$ one has type I superconductor,
while for $m_H>m_\gamma$ it is of type II, this is related to the fact
that scalar field produces an attraction for two vortices,
while the electromagnetic field produces a repulsion.
The boundary separating superconductors of
the I and II type corresponds to $m_H=m_\gamma$, i.e. to
special value of quartic coupling $\lambda$
\be
\lambda\ = \frac{n^2_eg^2}{2}  = {g^2\over 6}
\label{lambda}
\ee
This is exactly the value of quartic coupling in (\ref{ah}) we get
from the potential (\ref{pot})
using the ansatz (\ref{ans01}), (\ref{ansu}).
In this case vortices do not interact.
It is well known that vanishing of interaction when $m_H=m_\gamma$ can be
explained by the BPS nature of the ANO strings. The ANO string
satisfies the first order equations and saturate
the Bogomolny bound~\cite{B}, which can be found from the
following representation of string tension $T$ (see
(\ref{ah})),
\be
T =2\pi\zeta \,|n|+\int{d}^2
x\left\{\left(\frac1{2g}F_{IJ}\pm \frac{g}{2\sqrt{3}}
\left(|\varphi|^2-\zeta\right)\epsilon_{IJ}\right)^2
+ \frac12\left|\nabla_I \,\varphi\pm i\epsilon_{IJ}
\nabla_J\, \varphi\right|^2 \right\}
\label{tens}
\ee
Here indices $I,J=1,2$ denote coordinates transverse to the axis of the
vortex. For positive $n$ we take the upper sign in (\ref{tens}), whereas
for negative $n$ we take the lower sign.
The minimal value of the tension is reached when both positive
terms in the integrand of (\ref{tens}) vanish, then the string tension becomes
\be
 T_{BPS}\ =\ 2\pi\zeta  \,| n|
\label{tension}
\ee
where the winding number $n$ counts the magnetic flux $2\pi n$.
The linear dependence of string tensions on $n$ is consistent with
the absence of string interactions.

For the $n=1$ case vanishing of the integrand in (\ref{tens})
leads to well-known two first order differential equations
\be
r\frac{\rm d}{{\rm d}r}\,\phi (r)- f(r)\,\phi (r)\ =\ 0\
\\
-\frac1r\,\frac{\rm d}{{\rm d}r} f(r)+\frac{g^2}{3}\,
\left(\phi^2(r)-\zeta\right)\ =\ 0\
\label{foe}
\ee
(for the positive signs in (\ref{tens})),
where the profile functions $\phi(r)$ and $f(r)$ are introduced
in a standard way, i.e.
\be
\varphi(x) = \phi (r)\, {\rm e}^{i\,\vartheta}\
\\
\label{prof}
A^{(u)}_I(x) = \sqrt{3}\epsilon_{IJ}\,\frac{x_J}{r^2}\ [f(r)-1]
\ee
Here $r=\sqrt{\sum_{J=1,2} x_J^2}$ is the distance from
position of the vortex and $\vartheta$
($\d_I\vartheta = -\epsilon_{IJ}x_J/r^2$) is polar
angle in transverse to the axis of vortex
$(1,2)$-plane. The profile functions are real and
satisfy the boundary conditions
\be
 \phi (0)=0\ ,
\qquad ~f(0)=1\
\\
  \phi (\infty)=\sqrt{\zeta}\ , \quad
f(\infty)=0\
\label{bc}
\ee
which ensures that scalar field reaches its VEV $\sqrt{\zeta}$ at
infinity and vortex carries one unit of magnetic flux.
Equations (\ref{foe}) with boundary conditions (\ref{bc})
lead to unique solution for the profile functions
(although an analytic form of this solution is not found, though for $\zeta=0$ the
system (\ref{foe}) is equivalent to the "radial" Liouville equation).
The tension of string with winding number $n=1$ is given
by particular case of (\ref{tension})
\be
\label{r1ten}
T_{n=1} = 2\pi \zeta  = 6\pi|\mu m|
\ee
In \N2 QED  emergence
of the first order equations (\ref{foe}) means that some (half) of
the SUSY charges of \N2 algebra act trivially onto the ANO solution
(cf. \cite{HS,DDT,GS,VY}). In this case the Bogomolny
(topological) bound for the string tension coincides with the
central charge of SUSY algebra.

Now let us discuss the embedding of flux (winding) numbers of the string
(\ref{ans01}), (\ref{ansu}), (\ref{prof}) into the Cartan subalgebra of
$SU(3)$ group. Throughout  this paper we will use the convention of labeling
the flux of a given ANO string by magnetic charge of the monopole which
produces this flux and can be attached to its end \footnote{This monopole
can be a superposition of "really existing" monopoles in given vacuum as we
will see below.}, this is possible since both string fluxes and monopole
charges are elements of the group $\pi_1(U(1)^{\otimes 2}) =
{\bf Z}^{\otimes 2}$. This convention is convenient because specifying
the flux of given string we automatically fix the charge of monopole confined
by this string.

The string solution (\ref{ans01}), (\ref{ansu}), (\ref{prof}) has non-zero
$A^{(u)}_{\mu}$ while $A_{2\mu}$ vanishes, i.e. the matrix $A_{\mu}$ for this
configuration looks like
\be
\label{ust}
A_{\mu} = \frac{A^{(u)}_{\mu}}{\sqrt{3}}\left(
\begin{array}{ccc}
  1 & 0 & 0 \\
  0 &   -\frac12 & 0 \\
  0 & 0 & -\frac12
\end{array}\right)
\ee
It preserves the $SU(2)$ subgroup acting in the $2\times 2$ right low
corner of this matrix. Thus the flux of this string (charge of the
monopole attached to its end) is orthogonal to the root vector
$\balpha_2$ (\ref{e2m}) (or ${\bf e}_2$,
see fig.~\ref{fi:su3}). Hence, the string charge vector is proportional to
$u$-quark weight vector on the Cartan plane, ${\bmu}_u\sim \bmu_1$.
We call this string $u$-string and for this reason denote its charge vector
${\bf q}_{u}$.

To work out the  absolute value of the $u$-string charge we recall that the
Dirac quantization condition in the ${\tt r}=1$ vacuum with
non-zero VEV of the $u$-quark looks like
\be
\label{dirac}
{\bf q}_u{\bf u}={n\over 2}
\ee
Thus for winding number $n=1$ one has
\be
\label{1alpha}
{\bf q}_u=\frac32 {\bf u}
\ee
where we use that $|{\bf u}|=n_e=1/\sqrt{3}$, see (\ref{nabla}).

\begin{figure}[tb]
\epsfysize=8cm
\centerline{\epsfbox{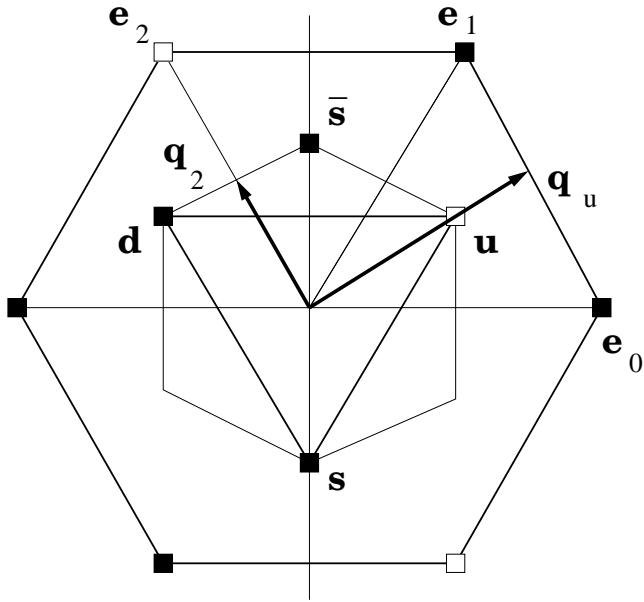}}
\caption{\sl Strings and different phases in ${\tt r}=1$ vacuum.
Black squares correspond to the monopole
and quark states in confinement phase while white squares -- to the states in Higgs
phase. Vectors ${\bf q}_u$ and ${\bf q}_2 = {\bf e}_2/2$ label the (charges of)
magnetic and electric strings correspondingly.}
\label{fi:phases}
\end{figure}

From (\ref{1alpha}) we see that absolute value of the string
charge is $|{\bf q}_u|=\sqrt{3}/2$ and this vector points into
the middle of the side of adjoint hexagon, see fig.~\ref{fi:phases}.
Note, that monopole
charges (whose positions in the Cartan plane coincide with the
positions of W-boson charges) correspond to the corners of
adjoint hexagon. We see that the component of monopole charge
parallel to vector ${\bf u}$ is confined by the ANO ${\bf u}$-string.
Moreover, it is easy to see from fig.~\ref{fi:phases} that the
values of projections of charges of ${\bf e}_1$ and ${\bf e}_0$
monopoles onto the ${\bf u}$-direction exactly matches the charge of
the ${\bf u}$-string. Thus ${\bf u}$-string confines ${\bf
u}$-component of  ${\bf e}_1$ and ${\bf e}_0$ monopole
charges.  This remark can be used as an explanation of rather
strange from group theory point of view value of charge of the
$u$-string lying outside the root lattice of $SU(3)$ group.

However, all monopoles present in the theory (${\bf e}_1$, ${\bf e}_2$ and
${\bf e}_0$- monopoles) have also non-zero component
in the orthogonal direction to vector ${\bf u}$, i.e. along ${\bf e}_2$.
What happens to this component of the monopole charge?
The answer to this question can be found in quantum theory, where the story
is a bit more complicated. As we already explained in sect.~\ref{ss:su3}
the $SU(2)$ gauge subgroup is preserved only classically at ${\tt r}=1$
vacua, and from the Seiberg-Witten exact solution \cite{SW1,SW2}
we know that in quantum theory $SU(2)$-subsector runs into strong
coupling regime where the $SU(2)$ subgroup is never restored;
instead, one gets either monopole or dyon vacuum.
Consider, say, the monopole one. At nonzero $\mu$ monopole
develops VEV of the order of $\mu \Lambda_{SU(2)}$ and this monopole
has charge ${\bf e}_2$, see (\ref{phisu2}). It means that
${\bf e}_2$-components of monopole charges are
screened by the condensation of the ${\bf e}_2$-monopole.

Passing to dual magnetic theory at strong coupling
one can also study the formation of the ANO {\em electric} string due
to condensation of
${\bf e}_2$-monopole along  same lines as above. It is clear from the Dirac
quantization condition that
${\bf e}_2$-string arises with the charge ${\bf q}_2={\bf e}_2/2$,
see fig.~\ref{fi:phases}.
It confines ${\bf e}_2$-components of $d$- and $s$-quark electric
charges while
their ${\bf u}$-component is screened by the $u$-quark condensation.
Note again, that ${\bf e}_2$-components of $d$- and $\bar{s}$-quark
electric charges exactly coincides with the charge of
the ${\bf e}_2$-string.
The electric ${\bf e}_2$-string which confines the
fundamental charge is very similar to strings at monopole vacua
of $SU(3)$ gauge theory without fundamental matter (labeled by
${\tt r}=0$ in our notations) studied in \cite{EFMG}.

To sum up, in each ${\tt r}=1$ vacuum we obtain one BPS magnetic ANO
${\bf u}$-string and one BPS electric ${\bf e}_2$-string.
In other words we have mixed Higgs phase for
$u$-quark and ${\bf e}_2$-monopole and confining phase for
${\bf e}_1$- and ${\bf e}_0$-monopole as well as for
$d$- and $s$-quark, see fig.~\ref{fi:phases}.

\subsection{Strings in r=2 vacua
\label{ss:r2strings}}

Now let us turn to strings in ${\tt r}=2$ vacua. For simplicity we
assume, first, the presence of only two quark flavors with masses $m_1$ and
$m_2$ under the same conditions as in sect.~\ref{ss:r2spec}, i.e. we
consider $s$-quark to be heavy and ignore it in low-energy theory.
The VEV's of all fields are given by (\ref{vacr2}) and (\ref{r2vev}),
in particular
VEV's of $u$- and $\tilde{u}$-quarks are equal to each other
as well as VEV's of $d$- and $\tilde{d}$-quarks. Therefore, we look
for string solutions using the following ansatz
\be
\label{1st}
u^{A}=\bar{\tilde{u}}_{A}= \delta_{A,1}\frac{\varphi_u}{\sqrt{2}}
\\
d^{A}=\bar{\tilde{d}}_{A}= \delta_{A,2}\frac{\varphi_d}{\sqrt{2}}
\ee
while the fields $a_3$ and $a_8$ in the leading order in $\mu/m$ are
given by their VEV's.

With this ansatz the effective action (\ref{qed}) becomes
\be
S=\int {\rm d}^4x\left(\frac1{4g^2}\left(F^{(3)}_{\mu\nu}\right)^{2}
+ \frac1{4g^2}\left(F^{(8)}_{\mu\nu}\right)^{2}
 +|\nabla^{(u)}_\mu \varphi_u|^2
+|\nabla^{(d)}_\mu \varphi_d|^2 +\right.
\\
\left. + \frac{g^2}{8}\left(
|\varphi_u|^2-|\varphi_d|^2-\xi (1-\omega)\right)^2 +
\frac{g^2}{24}\left(
|\varphi_u|^2+|\varphi_d|^2-\xi (1+\omega)\right)^2\right)
\label{r2qed}
\ee
where we use the same notations as in (\ref{nabla}) and (\ref{Aud}) while
$\xi \equiv |\varphi_u|^2$ and $\omega\equiv |\varphi_d|^2/|\varphi_u|^2$
are introduced in (\ref{xi}) and (\ref{x}) (note that
we consider both of
them  real and positive).  Gauge fields $A^{(u)}_{\mu}$ and
$A^{(d)}_{\mu}$ here as functions of orthogonal fields
$A^{(3)}_{\mu}$ and $A^{(8)}_{\mu}$  are given by (\ref{Aud}).
To study more general case of complex $\xi$ and
$\omega$ one should modify ansatz (\ref{1st}) taking into
account relative phases of the fields $Q$ and $\tilde{Q}$.

Let us now derive the Bogomolny bound for string solutions
in the theory (\ref{r2qed}).
Assuming that all fields depend only on two spatial coordinates
orthogonal to the string axis one can rewrite (\ref{r2qed}) as follows
\be
T =\int{d}^2 x\left(\left[\frac1{2g}F^{(3)}_{IJ} \pm
     \frac{g}{4}
\left(|\varphi_u|^2- |\varphi_d|^2-\xi (1-\omega)\right)
\epsilon_{IJ}\right]^2+
\right.
\\
+\left[\frac1{2g}F^{(8)}_{IJ} \pm
     \frac{g}{4\sqrt{3}}
\left(|\varphi_u|^2+ |\varphi_d|^2-\xi (1+\omega)\right)
\epsilon_{IJ}\right]^2+
\\
\left.+ \frac12\left|\nabla^{(u)}_I \,\varphi_u \pm i\epsilon_{IJ}
\nabla^{(u)}_J\, \varphi_u\right|^2
+ \frac12\left|\nabla^{(d)}_I \,\varphi_d \pm i\epsilon_{IJ}
\nabla^{(d)}_J\, \varphi_d\right|^2
\pm
\frac{1}{\sqrt{3}}\left(\tilde{F}^{(u)} +
 \tilde{F}^{(d)} \omega \right)\xi
\right)
\label{bog}
\ee
where we introduced (two-dimensional) dual field strength for gauge
fields $A^{(u,d)}_{\mu}$ as
$\tilde{F}^{(u,d)}=\frac12\epsilon_{IJ}F_{IJ}^{(u,d)}$.
The upper sign here corresponds to positive total flux,
the value of $\frac{1}{\sqrt{3}}\int d^2x\left(\tilde{F}^{(u)} +
\tilde{F}^{(d)}\omega\right)$, while
the lower sign corresponds to the negative total flux.

 The last term in (\ref{bog}) represents exactly the string fluxes
while all other
positive terms in (\ref{bog}) should be zero on the BPS solutions, leading
to the following first order equations
\be
\label{F3d}
\frac1{2g}F^{(3)}_{IJ}+
     \frac{g}{4}\varepsilon
\left(|\varphi_u|^2- |\varphi_d|^2-\xi (1-\omega)\right)\epsilon_{IJ}=0
\\
\frac1{2g}F^{(8)}_{IJ}+
     \frac{g}{4\sqrt{3}}\varepsilon
\left(|\varphi_u|^2+ |\varphi_d|^2-\xi (1+\omega)\right)\epsilon_{IJ}=0
\\
\nabla^{(u)}_I \,\varphi_u+i \varepsilon\epsilon_{IJ}
\nabla^{(u)}_J\, \varphi_u=0
\\
\nabla^{(d)}_I \,\varphi_d+i\varepsilon\epsilon_{IJ}
\nabla^{(d)}_J\, \varphi_d=0
\ee
where $\varepsilon = \pm$ is the sign of total flux.

One can classify possible solutions to these equations
by behavior of the fields at spatial infinity in $(1,2)$ plane.
One type of solutions has nontrivial winding for the $u$-quark another
type has winding for the $d$-quark and there are also mixed solutions
when both $u$- and $d$-quarks wind at infinity.

Let us start with string solutions when only $u$-quark winds.
Assuming for simplicity the unit
winding number one has at $x\to \infty$
\be
\varphi_u\sim e^{i\vartheta}\sqrt{\xi}
\\
\varphi_d\sim \sqrt{\omega\xi}
\label{1ud}
\ee
This behavior requires the following behavior at infinity
for the gauge fields
\be
\label{1Au}
\frac1{\sqrt{3}}A^{(u)}_{I}\stackreb{r\to\infty}{\to} \partial_I\vartheta
\ee
and
\be
\label{1Ad}
A^{(d)}_{I}\stackreb{r\to\infty}{\to} 0
\ee
which ensures finite contribution to string tension
coming from kinetic terms of $u$- and $d$-quarks.
Such behavior at infinity means that the flux
of the field $A^{(d)}_{\mu}$ vanishes while the flux
of the $A^{(u)}_{\mu}$ field is
\be
\frac1{\sqrt{3}}\int d^2 x \tilde{F}^{(u)} = 2\pi
\ee
for unit winding number, then
representation (\ref{bog}) for the BPS string tension gives
\be
\label{1ten}
T_1=\,2\pi \xi\,=\,4\pi\, \left|\mu(2m_1+m_2)\right|
\ee
To implement conditions (\ref{1Ad}) let us rotate the gauge fields
$A^{(3)}_{\mu}$, $A^{(8)}_{\mu}$ to another orthogonal combinations
$A^{(d)}_{\mu}$ (given by (\ref{Aud})) and $A_{1\mu}$ (the component
along the vector ${\bf e}_1$) as
\be
\label{A1}
A_{1\mu}=\frac12 A^{(3)}_{\mu}+\frac{\sqrt{3}}{2}A^{(8)}_{\mu}
\\
A^{(d)}_{\mu}=-\frac{\sqrt{3}}{2}A^{(3)}_{\mu}+\frac12 A^{(8)}_{\mu}
\ee
and rewrite our first order eqs. (\ref{F3d}) in terms of
these fields.

The behavior at infinity (\ref{1ud})-(\ref{1Ad}) suggests that one can look
for a solution in terms of the following profile functions
\be
\label{1prof}
\varphi_u(x) = \phi_u (r) {\rm e}^{i\,\vartheta},\;
\varphi_d (x) = \phi_d (r)
\\
A_{1I}(x) = 2\epsilon_{IJ}\,\frac{x_J}{r^2}\ [f_1(r)-1],\;
A^{(d)}_I(x) =\sqrt{3}\epsilon_{IJ}\,\frac{x_J}{r^2}\ f_d(r)
\ee
Note, that the charge of $u$-quark with respect to $A_{1\mu}$ field
is $n_e=1/2$, while the charge of $d$-quark with respect to $A^{(d)}_{\mu}$
is $n_e=1/\sqrt{3}$, which exactly corresponds to the coefficients in
(\ref{1prof}).
With this substitution the first order equations (\ref{F3d}) turn into the
system of four first-order nonlinear differential equations
\be
r\frac{\rm d}{{\rm d}r}\,\phi_u (r)- f_1(r)\,\phi_u (r)\
+ \frac12 f_d(r)\,\phi_u (r) =\ 0
\\
r\frac{\rm d}{{\rm d}r}\,\phi_d (r)- f_d(r)\,\phi_d (r)\ =0
\\
-\frac1r\,\frac{\rm d}{{\rm d}r} f_1(r)+\frac{g^2}{4}\,
\left(\phi_u(r)^2-\xi\right)\ =\ 0
\\
 -\frac1r\,\frac{\rm d}{{\rm d}r} f_d(r)+\frac{g^2}{3}\,
\left(\phi_d(r)^2-\frac12\phi_u(r)^2+\frac12\xi-\omega\xi\right)\ =\ 0
\label{1foe}
\ee
The reason why the fields $\varphi_d-\sqrt{\omega\xi}$
and $A^{(d)}_{\mu}$ with trivial
behavior at infinity cannot be simply put to zero is mixing between
$\varphi_u$, $A_{1\mu}$ and $\varphi_d$, $A^{(d)}_{\mu}$ since
the fields $\varphi_u$ and $A_{1\mu}$ which wind at infinity
are not eigenvectors of the mass matrix (\ref{r2phmat}). In other
words
(\ref{1foe}) does not have nontrivial solutions with $f_d\equiv
0$ and $\phi_d\equiv \sqrt{\omega\xi}$
(which is a naive ansatz suggested by  the
boundary conditions (\ref{1ud}), (\ref{1Ad})).

First order equations (\ref{1foe}) should be supplemented by the
boundary conditions
which ensure that quark fields go to their VEV's at infinity,
$u$-quark field has no singularity at $r=0$ ($\phi_u (0)=0$) and $A_{1\mu}$ field
carries flux equal to $2\pi$ ($f_1(0)=1$).

\begin{figure}[tb]
\epsfysize=9cm
\centerline{\epsfbox{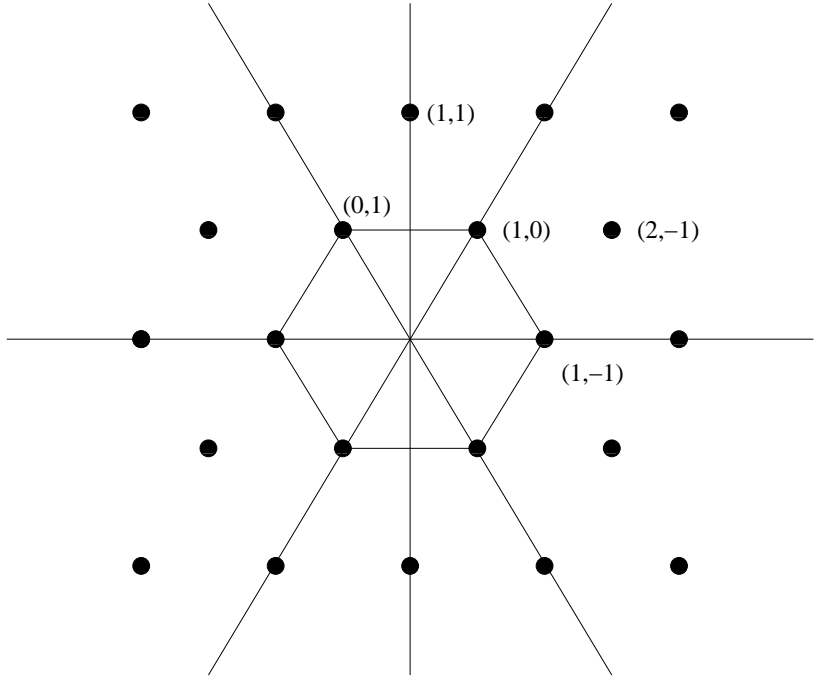}}
\caption{\sl Lattice of string solutions in ${\tt r}=2$ vacuum. We have specified explicitly
$(1,0)={\bf e}_1$, $(0,1)={\bf e}_2$, $(1,-1)={\bf e}_0$,
$(1,1)={\bf e}_1+{\bf e}_2$ and $(2,-1)\equiv 2{\bf u}$ strings.}
\label{fi:lattice}
\end{figure}

Now let us discuss the charges of our string. Since $A^{(d)}_{\mu}$ goes to
zero at infinity the gauge field matrix looks like
\be
\label{1stA}
A_{\mu} \stackreb{r\to\infty}{\rightarrow} \frac{A_{1\mu}}{2}\left(
\begin{array}{ccc}
  1 & 0 & 0 \\
  0 &   0 & 0 \\
  0 & 0 & -1
\end{array}\right)
\ee
at infinity.
Eq.~(\ref{1stA}) means that the charge of string is directed
along the root $\balpha_{12}\sim {\bf e}_1$ in the Cartan algebra.
Its charge vector should satisfy the Dirac quantization condition
${\bf q}_1{\bf u}=1/2$ ($u$-quark winds) which gives
\be
\label{1str}
{\bf q}_1 = {\bf e}_1
\ee
This is the reason why we call this string ${\bf e}_1$-string. Its charge
is directed to the upper right corner of the adjoint hexagon, see
fig.~\ref{fi:lattice}.

In a similar way one can consider the string solution when $d$-quark
winds at infinity, while $u$-quark field runs to its VEV.
This gives us ${\bf e}_2$-string with charge vector directed to the
upper left corner of the adjoint hexagon, see fig.~\ref{fi:lattice}.
The easiest way to describe this string is to use another set of
orthogonal gauge potentials $A_{2\mu}$ and $A^{(u)}_{\mu}$ related to
$A_{1\mu}$ and $A^{(d)}_{\mu}$ by
\be
\label{Au2}
A^{(u)}_{\mu}=\frac{\sqrt{3}}{2}A_{1\mu}-\frac12 A^{(d)}_{\mu}
\\
A_{2\mu}= \frac12 A_{1\mu} +\frac{\sqrt{3}}{2}A^{(d)}_{\mu}
\ee
The behavior at infinity of ${\bf e}_2$-string is given
by
\be
\label{2prof}
\varphi_d(x) = \phi_d (r)\, {\rm e}^{i\,\vartheta},\;
\varphi_u (x)=\phi_u (r)
\\
A_{2I}(x) = 2\epsilon_{IJ}\,\frac{x_J}{r^2}\left(f_2(r)-1\right),\;
A^{(u)}_I(x) = \sqrt{3}\epsilon_{IJ}\,\frac{x_J}{r^2}\ f_u(r)
\ee
The profile functions here satisfy equations and
boundary conditions similar to (\ref{1foe}).
The tension of the ${\bf e}_2$-string is given by
\be
\label{2ten}
T_2=\,2\pi \omega \xi \,=\,4\pi\,\left|\mu (2m_2+m_1)\right|
\ee
and this string, if exists, is also BPS saturated.

In general, one can also consider string solutions when both
$u$- and $d$-quarks wind. Let us fix some basis, say, the
potentials $A_{1\mu}$ and $A^{(d)}_{\mu}$ (\ref{A1}) and introduce the profile
functions for the string when $u$-quark winds $n$ times
and $d$-quark winds $k$ times (to be denoted as $(n,k)$-string
since its charge
in our normalization is ${\bf q}_{n,k}=n{\bf e}_1 + k{\bf e}_2$)
defined as
\be
\label{nmpr}
\varphi_u(x) = \phi_u (r)\, {\rm e}^{in\vartheta},\;
\varphi_d (x)=\phi_d (r){\rm e}^{ik\vartheta},
\\
A_{1I}(x) = 2\varepsilon\epsilon_{IJ}\,\frac{x_J}{r^2}\
\left(f_1(r)-f_1(0)\right),\;
A^{(d)}_I(x) = \sqrt{3}\ \varepsilon\epsilon_{IJ}\,\frac{x_J}{r^2}\
\left(f_d(r)- f_d(0)\right)\
\ee
These profile functions satisfy the first order equations (\ref{1foe}),
the only modification is due to the extra sign factor $\varepsilon$ in
(\ref{nmpr}) which accounts for
possible positive or negative total flux
\be
\label{sign}
\varepsilon = \varepsilon_{n,k} = \frac{n+\omega k}{|n+\omega k|} =
{\rm sign}(n+\omega k) = \pm 1
\ee
The boundary conditions for profile functions of $(n,k)$-string read
\be
\label{phibc}
\phi_u (0)=0,\; \phi_d (0)=0,\;
\phi_u (\infty)=\sqrt{\xi},\; \phi_d (\infty)=\sqrt{\omega\xi}
\ee
for the scalar fields and
\be
f_1(0) = \varepsilon_{n,k}\left( n+\frac{k}{2}\right),
\;f_d(0)=\varepsilon_{n,k}\,k, \;
f_1(\infty)=0, \;f_d(\infty)=0
\label{fbc}
\ee
for the gauge fields. The boundary condition for $f_1(0)$ is easy to derive
from $f_u(0)=\varepsilon_{n,k}\,n$ using relations (\ref{Au2}).

From the boundary conditions (\ref{fbc}) and two first equations of
(\ref{1foe}) one immediately finds the leading behaivior of the
profile functions at $r\to 0$
\be
\label{fir0}
\phi_u(r) \stackreb{r\to 0}{\propto} r^{f_1 - f_d/2} = r^{\epsilon_{n,k}n}
\\
\phi_d(r) \stackreb{r\to 0}{\propto} r^{f_d} = r^{\epsilon_{n,k}k}
\ee
which means that nonsingular at the origin solutions to the BPS first-order
equation exist only for both $n$ and $k$ either positive or negative
simultaneously (dependently on the sign of $\epsilon_{n,k}$).

The tension of $(n,k)$-string is given again by last term in the
representation (\ref{bog}) and reads
\be
\label{nmten}
T_{n,k}=\,2\pi \xi |n+\omega k| =
4\pi\left|\mu \left(
n(2m_1+m_2)+ k(2m_2+m_1)\right)\right|
\ee
The lattice of possible BPS $(n,k)$-strings is shown in
fig.~\ref{fi:lattice}; ${\bf e}_1$-string is identical to $(1,0)$-string
while ${\bf e}_2$-string is nothing but $(0,1)$-string. There is one
more string whose charge belongs to the
adjoint hexagon, namely the $(1,-1)$-string (to be called
also ${\bf e}_0$-string in what follows), see fig.~\ref{fi:lattice}.
As is clear from fig.~\ref{fi:lattice},
${\bf e}_1$-string can be considered as a bound state of
${\bf e}_0$ and ${\bf e}_2$-string, which exactly corresponds to
decomposition of the root $\balpha_{12}$ into linear combination of
the simple roots $\balpha_1$ and $\balpha_2$, see fig.~\ref{fi:su3}.\
It is clear, thus, that string charges are labeled by (normalized to unity)
root vectors of the $SU(3)$ root lattice. We will return to this question in
sect.~\ref{ss:strings}, where we study
which of the BPS $(n,k)$-strings exist as
solutions to the first order equations (\ref{1foe}) and which among them are
stable. Here we conclude by considering certain
special cases in which some of $(n,k)$-strings becomes ANO strings
and equations (\ref{1foe}) reduce to
ANO equations (\ref{foe})  with one gauge and one scalar potential.

Consider, first, the simplest example of this kind with both
quarks having unit winding number, namely the $(1,1)$- or
$({\bf e}_1+{\bf e}_2)$-string. This string can be considered as
bound state of ${\bf e}_1$- and ${\bf e}_2$-strings. If this bound
state exists and saturates the Bogomolny bound
then (\ref{nmten}) suggests for its tension
\be
\label{12ten}
T_{1,1}=\,12\pi\,\left|\mu (m_1+m_2)\right|
\ee
Now let us give some evidence that this BPS string indeed exists.
Consider the special value of parameter (\ref{x}) $\omega=1$, then VEV's
of $u$- and $d$-quarks are equal. The string solution can be easily
constructed with the only component of gauge field $A^{(8)}_{\mu}$
being non-zero (putting $A^{(3)}_{\mu}=0$)
and one complex scalar field introduced via
\be
\label{phi12}
\varphi_u=\varphi_d=\frac{\varphi}{\sqrt{2}}
\ee
With this ansatz the theory (\ref{r2qed}) reduces to standard
Abelian Higgs model of type (\ref{ah}) with equal values for photon and
scalar masses $m_{\gamma}=m_H=g^2\xi/3$ and
this model obviously possesses the BPS ANO string solution.
This mass of the Abelian Higgs model coincides with
the eigenvalue of the photon mass matrix, corresponding to $\Omega =2/3$,
see (\ref{r2mass}), (\ref{lam}). This is one of possible explanations
why a single gauge potential and single scalar field
appear in the string solution at $\omega=1$:
gauge field $A^{(8)}_{\mu}$ and scalar (\ref{phi12})
correspond to eigenvectors of gauge and scalar mass matrices.

One can reach the same conclusion directly from equations (\ref{1foe}).
Substituting the ansatz $\phi_u=\phi_d$ and $f_1=3/2\;f_d$ (this is
consistent with boundary conditions (\ref{fbc}) because $f_1(0)=3/2$
and $f_d(0)=1$ for $(1,1)$-string) we see that four equations (\ref{1foe})
at $\omega=1$ reduce exactly to the system of two ANO equations (\ref{foe}),
under the following  modifications: the ${\tt r}=1$ parameter $\zeta$ is replaced by
the ${\tt r}=2$ parameter $\xi$ and the coefficient in front
of the second term in the last equation of (\ref{foe}) replaced
by $m_{\gamma}^2/2\xi=g^2/6$ for $\omega=1$.

Now let us move parameter $\omega$ away from $\omega=1$. It is clear that
under continuous deformation string solution cannot disappear
or become non-BPS state, since BPS string belongs to a
short multiplet which cannot turn into long one without breaking of
some amount of supersymmetry (the number of states cannot jump).
Thus, at least at some region of values of parameter $\omega$
we expect existence of the BPS $({\bf e}_1+{\bf e}_2)$-string, we specify
this region in sect.~\ref{ss:lattice}.
where more detailed analysis of the nature of $(n,k)$-strings
and their stability, based on study of
string interactions, will be presented.
One may also use more direct evidence based on numerical simulations
(we have done this using the MAPLE program) of the BPS first order
equations (\ref{1foe})
\footnote{We tried to use the string solution
in  monopole vacuum presented in \cite{EFMG}, but unfortunately it does
not seem to satisfy all four eqs.(\ref{1foe}).}.

To conclude this section note, that electric flux tubes
quite similar to our magnetic $(n,k)$-strings were found
in \N2 $SU(3)$ theory without fundamental matter
at strong coupling vacua \cite{EFMG}.
These strings appear due to the condensation of monopoles/dyons
and their charges
should be associated with the (normalized) weight lattice on the Cartan
plane instead of (normalized) root lattice which arises for our magnetic
strings.  Strings of this type are often called $Z_3$ strings
when studied directly in the underlying non-Abelian $SU(3)$
theory, see  \cite{V,VS,HV,SS,KB,KS}.

\subsection{BPS formula for the string tensions}

In this section we rewrite the mass formulas for the BPS strings
studied in previous sections in a form which is more
familiar for the BPS objects.
Tensions of BPS strings are given by central charges of \N2
SUSY algebra \cite{HS,DDT,GS,VY}.
They have general structure
\be
\label{bpsten}
T_{BPS}=2\pi |{\bf q}_{s}{\bf f}|
\ee
where, in our particular case, ${\bf q}_{s}$ is the charge vector of a given string
in the Cartan plane (see fig.~\ref{fi:phases} and fig.~\ref{fi:lattice})
and ${\bf f}$ is the generalized vector parameter
of the FI $F$-term, which can be defined as
\be
\label{vxi}
{\bf f} = -4\mu \bphi
\ee
where (see (\ref{phigen}), (\ref{e1m}) and (\ref{e2m}))
\be
\label{vecphi}
\bphi= \phi_1 {\balpha}_{12} +\phi_2 {\balpha}_2
= \sqrt{2}(\phi_1{\bf e}_1+ \phi_2{\bf e}_2)
\ee
where $\phi_1$ and $\phi_2$ are the corresponding components of
the VEV's of the adjoint scalar matrix (\ref{diag}) in
given vacuum. In the orthogonal
basis of fields $a_3$ and $a_8$ they are equal to (see (\ref{phigen}))
\be
\label{a12}
\phi_1 = \2\left(a_3+\frac{a_8}{\sqrt{3}}\right), \;\;
\phi_2 = \2\left(-a_3+\frac{a_8}{\sqrt{3}}\right)
\ee
Let us now show that (\ref{bpsten}) indeed gives correct tensions for all
BPS strings considered before. Start with ${\tt r}=1$ vacua,
from (\ref{phir1}) one finds that
\be
\label{1phi}
\bphi = -m\left({\bf e}_1 - {{\bf e}_2\over 2}\right)
\ee
In  ${\tt r}=1$ vacuum we have only ${\bf u}$-string with the charge
${3\over 2}{\bf u}$,
see (\ref{1alpha}). For this particular string (\ref{bpsten}) gives
\be
\label{t1}
T_{u}=8\pi \left|\frac32\mu m\ {\bf u}
\left({\bf e}_1 - {{\bf e}_2\over 2}\right)\right|=6\pi|\mu m|
\ee
which coincides with (\ref{r1ten}). Now consider ${\tt r}=2$ vacuum,
where one has (see (\ref{r2phi}))
\be
\label{2phi}
\bphi = - m_1 {\bf e}_1 -  m_2 {\bf e}_2
\ee
thus, for the $(n,k)$-string formula (\ref{bpsten}) gives
\be
T_{n,k}=8\pi |\mu (n{\bf e}_1+  k{\bf e}_2)
(m_1{\bf e}_1 +  m_2{\bf e}_2)|=4\pi\left|\mu \left(
n(2m_1+m_2)+ k(2m_2+m_1)\right)\right|
\ee
and this result coincides with (\ref{nmten}).
The BPS formula (\ref{bpsten}) is valid in the limit
$\mu \to 0$, $\xi\sim \mu m_A =const$, when one can neglect
breaking of \N2 SUSY in effective QED (\ref{qed}) and it assumes also the
weak coupling regime $|m_A|\gg\Lambda$.

We conclude this section noting that masses of W-bosons are also
determined by the BPS formula, following from (\ref{Higgs})
\be
\label{wbps}
m_{W} = \sqrt{2}|{\balpha}_{W}{\bphi}|=
2|{\bf q}_{W}{\bphi}|
\ee
where ${\bf q}_{W}$ is the (normalized) charge of corresponding W-boson in the
Cartan plane. In particular, it means that tensions of ${\bf e}_0$-,
${\bf e}_1$-
and ${\bf e}_2$-strings which belong to the adjoint hexagon are
proportional to the
masses of corresponding W-bosons at least at weak coupling
when $m_{W}\gg\Lambda$. Indeed, charge vectors ${\bf q}_s$
of the ${\bf e}_0$-, ${\bf e}_1$- and ${\bf e}_2$-strings coincide with
the charges ${\bf q}_{W}$ of related W-bosons, and comparing
(\ref{bpsten}) with (\ref{wbps}) we see that tensions of all strings are
proportional to the masses of corresponding W-bosons.

\section{(n,k)-strings
\label{ss:strings}}

In this section we address the question of existence and stability
of $(n,k)$-strings in ${\tt r}=2$ vacua studying their interactions
at large distances. This allows us to see which strings attract each
other and form stable bound states and which do not interact so that
their bound state is only marginally stable. First, we consider the
interaction of the ANO strings in order to develop necessary technique
and then turn to the interactions of $(n,k)$-strings.

\subsection{Interactions of ANO strings
\label{ss:intano}}

Let us first develop the method of effective vertex to
calculate the interactions of ANO strings. This method is well-known
in instanton physics and is used there to calculate instanton interactions
\cite{CDG,SVZ,Y90,Y96}. Here we
generalize it to the case of solitonic ANO strings (it also can be easily
generalized to any solitonic branes).

Transform the scalar and gauge fields of the ANO string from the
"regular" into "singular" gauge making the $U(1)$ gauge
transformation with $\exp\left(-i\vartheta\right)$,
where $\vartheta$ is polar angle in the $(1,2)$ plane orthogonal to
the string. This gauge transformation
is singular at $r=0$ so now the topological charge of the string
(flux) comes from small circle around the origin instead of
the large circle at $r\to\infty$, and in this gauge the substitution
(\ref{prof}) turns into
\be
\label{anosing}
\varphi(x) = \phi (r)\
\\
A_I(x) = \frac{\varepsilon_n}{n_e}
\epsilon_{IJ}\,\frac{x_J}{r^2}\ f(r)\
\ee
where the dependence on arbitrary electric charge $n_e$ is restored.
Here profile functions $\phi(r)$ and $f(r)$ satisfy the ANO
first order equations (cf. with (\ref{foe}))
\be
r\frac{\rm d}{{\rm d}r}\,\phi (r)-  f(r)\phi (r) = 0
\\
 -\frac1r\,\frac{\rm d}{{\rm d}r} f(r)+n_e^2 g^2
\left(\phi(r)^2-\xi\right) = 0
\label{gano}
\ee
and boundary conditions
\be
\phi (0)=0,
\quad f(0)= |n|
\\
\phi (\infty)=\sqrt{\xi}, \quad
f(\infty)=0
\label{anobc}
\ee
were $n$ is integer winding number, while
\be
\label{anosign}
\varepsilon_n =\frac{n}{|n|} = {\rm sign}(n)
\ee
Using equations (\ref{gano}) which determine the
exponential fall-off of the functions $f$ and $\phi-\sqrt{\xi}$ at infinity,
corresponding to the Yukawa behavior in two transverse dimensions,
we get for the large $r$ asymptotic
\be
\varphi (x)=\sqrt{\xi}\left(1-\frac{c_n}{\sqrt{m_{\gamma}r}}e^{-m_{\gamma}r}
+\cdots \right)
\\
A_I(x) = \frac{\varepsilon_n c_n}{n_e}\epsilon_{IJ}\,\frac{x_J}{r^2}
\sqrt{m_{\gamma}r}\;e^{-m_{\gamma}r}+\cdots
\label{anoinf}
\ee
where $c_n$ is the coefficient to be fixed below, while
$\varepsilon_n$ is given by (\ref{anosign}), both
$c_n$ and $\varepsilon_n $ depend on the winding number $n$.
We will also need the behavior of
the (two-dimensional) dual field strength $\tilde{F}=\frac12\epsilon_{IJ}
F_{IJ}$
at the infinity in two-dimensional plane $(1,2)$, and from (\ref{anoinf}) one
finds
\be
\label{anoinF}
\tilde{F}=\frac{\varepsilon_n c_nm_{\gamma}^2}{n_e}\ \frac{e^{-m_{\gamma}r}}
{\sqrt{m_{\gamma}r}}+\cdots
\ee
Now let us work out the effective vertex for the ANO string. This vertex
once added to the tree level QED action (e.g. (\ref{ah})) should reproduce
all effects of presence of the ANO string in the framework of
perturbation theory (see \cite{CDG,SVZ,Y90,Y96}).
For the string with winding number $n$ we propose the following form
\footnote{In what follows we are going to keep only bosonic background fields
in our expressions for the string interaction, the fermionic terms can be restored by
supersymmetry, similar to the instantonic case, see for example \cite{Y96}.}
(up to the overall normalization factor)
\be
V_{n}^{ANO}= \int DX(\bsigma)\exp \left\{-\int d^2\sigma 2\sqrt{2\pi}
 c_n \left[\bar{\varphi}\varphi(X(\bsigma)) +\frac{\varepsilon_n}
{2n_e g^2}F_{\mu\nu}(X(\bsigma))n_{\mu\nu}\right]\right\}
\label{anoV}
\ee
where fields $\varphi$ and $F_{\mu\nu}$ are considered as
functions of string co-ordinates $X(\bsigma)$. We also introduced here an
antisymmetric tensor $n_{\mu\nu}(X(\bsigma))$ orthogonal to the
string world-sheet at the point $X(\bsigma)$. For the case of
the straight string at rest directed along the third axis
$n_{\mu\nu}=0$ for $\mu$ or $\nu=0,3$, while $n_{IJ}=\epsilon_{IJ}$ for
$\mu,\nu=I,J=1,2$. Note, that here we use static parameterization
for the string world-sheet, $\sigma_1=t$ and $\sigma_2=x_3$, in other words
\be
\label{nmn}
n_{\mu\nu} = \frac12\epsilon_{IJ}{\d X_\alpha\over\d \sigma_I}
{\d X_\beta\over\d \sigma_J}
\epsilon_{\mu\nu\alpha\beta}
\ee
and this interaction is nothing but well-known interaction of string with
antisymmetric tensor $B$-field $B = *F$.

To check the expression for the effective vertex (\ref{anoV}) let us calculate
the correlation function
\be
\label{cor}
\langle \varphi(x)\ldots\tilde{F}(y)\ldots\rangle_{\rm string}
\ee
in the string background assuming that all points $x$'s and $y$'s
are far from the axis of the string at $X(\bsigma)$.
On one hand this correlation function can be calculated in semiclassical
approximation just substituting classical expressions (\ref{anoinf})
for the scalar and gauge fields at large distances from the string axis
into the correlation function (\ref{cor}),
on the other hand the same correlation function can be calculated
in perturbation theory as
\be
\label{Vcor}
\langle\varphi(x)\ldots\tilde{F}(y)\ldots
V_{n}^{ANO}\rangle
\ee
once the effective vertex $V_{n}^{ANO}$ is added to the tree level QED
action, and this should give rise to
the same result as a substitution of classical fields.

To see this expand scalar field around its VEV
$\varphi=\sqrt{\xi}+\delta\varphi$ and write down the bilinear in
scalar fields term in the exponential in (\ref{anoV}) as
\be
\label{scexp}
\bar{\varphi}\varphi =\xi +\sqrt{\xi}\delta\bar{\varphi}+
\sqrt{\xi}\delta\varphi
+ \dots
\ee
Consider, first, the linear in quantum fluctuations terms.
Expanding the exponential in (\ref{Vcor}) in
$\delta\bar{\varphi}$ and in $\tilde{F}$ we calculate the correlation
function (\ref{Vcor})
to the leading order in coupling constant using the
tree level propagator
\be
\label{sprop}
\langle \delta\varphi(x)\int d^2\sigma \overline{\delta\varphi}
(X(\bsigma))\rangle =
\frac{1}{2\pi}K_0 (m_{\gamma}r)\stackreb{r\to\infty}{\rightarrow}
\frac{1}{2\sqrt{2\pi}}\frac{e^{-m_{\gamma}r}}{\sqrt{m_{\gamma}r}}
+ \dots
\ee
for the massive scalar and
\be
\label{gprop}
\langle \tilde{F}(x)\int d^2\sigma \tilde{F}(X(\bsigma))\rangle =
- \frac{g^2}{2\pi}\frac{\partial^2}{\partial r^2}K_0 (m_{\gamma}r)
\stackreb{r\to\infty}{\rightarrow}
 -\frac{g^2 m_{\gamma}^2}{2\sqrt{2\pi}}
\frac{e^{-m_{\gamma}r}}{\sqrt{m_{\gamma}r}} + \dots
\ee
for the massive vector field, where $r$ is the distance between point $x$
and position of string in $(1,2)$ plane. These correlation functions are
nothing but propagators of two-dimensional theory rewritten
in four dimensional notations.

The quadratic in quantum fluctuations term in (\ref{scexp}) cannot be
verified using just tree level approximation (\ref{sprop}),
(\ref{gprop}), since it is next to leading effect in coupling constant.
Still one can restore the quadratic dependence on
field $\varphi$
in the exponential in (\ref{anoV}) observing that full effective
vertex can depend only on fields  ${\bar\varphi}$ and $\varphi$
rather than upon their VEV's. The result
looks a bit surprising from the point of view of string
theory, and the origin of this quadratic dependence is that
in our picture string tension
$T=(2\pi\alpha')^{-1}$ is "dynamical" and
determined, in contrast to the fundamental string theory, by
condensate of a scalar field.

Let us now determine the constant $c_n$. Ignoring quantum fluctuations
in (\ref{anoV}) and substituting the scalar field by its VEV $\sqrt{\xi}$
one gets $2\sqrt{2\pi}c_n \xi$ in the exponential in (\ref{anoV}), which
should be equal to string tension $2\pi\xi |n|$ (cf. (\ref{tension}));
that gives
\be
\label{an}
c_n=\sqrt{\frac{\pi}{2}}|n|
\ee
Substituting this into (\ref{anoV}) we finally obtain
\be
\label{anoVf}
V_{n}^{ANO}= \int DX(\bsigma)\exp \left\{-2\pi |n|
\int d^2\sigma   \left[\bar{\varphi}\varphi(X(\bsigma)) +\frac{\varepsilon_n}
{2n_e g^2}F_{\mu\nu}(X(\bsigma))n_{\mu\nu}\right]\right\}
\ee
and the effective partition function of low-energy QED is now given by
\be
\label{prtfun}
Z = \int DA_{\mu}D{\bar\varphi} D\varphi
\exp\left(-S_{QED}-\sum_{n}V_{n}^{ANO}\right)
\ee
The effective vertex in the exponent allows to have many strings
with different winding numbers, hence theory in (\ref{prtfun}) should
be treated perturbatively since non-perturbative effects (strings)
are already taken into account explicitly.

Let us now use the effective vertex (\ref{anoVf}) to compute the
interaction potential of two straight static strings directed along
the third axis at large separation $R$ in $(1,2)$ plane.
Expanding $\exp\left(-\sum_{n}V_{n}^{ANO}\right)$ in powers of $V_{n}^{ANO}$
we keep only the term $ V_{n_1}^{ANO}V_{n_2}^{ANO}$, related
directly to the interaction potential of strings $U_{1,2}$ via
\be
\label{defU}
\langle V_{n_1}^{ANO}V_{n_2}^{ANO}\rangle =
\int DX_1(\bsigma) DX_2(\bsigma)e^{-U_{1,2}V_{2}}
\ee
where $X_1(\bsigma)$ and $X_2(\bsigma)$ correspond to two strings while
$V_2$ is the volume of the two dimensional space in $(0,3)$-plane.
Using the propagators (\ref{sprop}) and (\ref{gprop}) to calculate
the correlation function in the l.h.s. of (\ref{defU}) we finally
obtain
\be
\label{anoU}
U_{1,2}=-(2\pi)^{\frac32}\xi |n_1||n_2|
\frac{e^{-m_{\gamma}R}}{\sqrt{m_{\gamma}R}}
\left[ 1-\varepsilon_1 \varepsilon_2 \right]
\ee
where  $\varepsilon_1$, $\varepsilon_2 $ refers to the signs of
winding numbers (\ref{anosign}) of two strings.

We see that potential of string interactions has exponential fall-off
at large separations $R$. The first term in square brackets comes from
the exchange by scalar field, and it always gives the attractive contribution
to the potential. The second term comes
from the photon exchange and its sign depends on the relative signs of
$n_1$ and $n_2$. If $n_1 n_2>0$ this term gives repulsion which cancels
attraction produced by scalar exchange, which is, of course, a
well known result -- the BPS ANO strings do not interact.
If, however, $n_1 n_2<0$ (which corresponds to string-antistring interactions)
the photon exchange also gives an attraction. Thus the total string-antistring
interaction potential is always attractive.

Let us conclude this section presenting the effective vertex of the ANO
string which takes into account its motion. It is clear that such
generalization of (\ref{anoVf}) has the form
\be
\label{string}
V_{n}^{ANO}= \int DX(\bsigma)\exp \left(-S^{\rm string}_n\right)
\\
S^{\rm string}_n= 2\pi |n|\int d^2\sigma \left[\sqrt{\det g_{\rm ind}}\
  \bar{\varphi}\varphi(X(\bsigma)) +\frac{\varepsilon_n}
{2n_e g^2}F_{\mu\nu}(X(\bsigma))n_{\mu\nu}\right]
\ee
Here $g_{ind}$ is the determinant of the induced metric given by
\be
\label{met}
g_{IJ}^{\rm ind}=\frac{\partial X_{\mu}}{\partial \sigma_{I}}
\frac{\partial X_{\mu}}{\partial \sigma_{J}}
\ee
and interaction with $B$-field is defined in (\ref{nmn}).
Besides the usual Nambu-Goto term the string action contains also
higher derivative terms omitted in (\ref{string}). The
effective action (\ref{string}) takes into account interactions of the bulk
fields $A_{\mu}$ and $\varphi$ with the two-dimensional "field"
$X_{\mu}(\bsigma)$ living on string world sheet.

\subsection{Interactions of (n,k)-strings}

In this section we apply the effective vertex method to calculate the
interaction potential of $(n,k)$-strings in ${\tt r}=2$ vacua.
First, from the first order equations (\ref{1foe})
we get the behavior of the fields $\phi_u(r)$, $\phi_d(r)$ and
$f_1(r)$, $f_d(r)$ at infinity
\be
\label{nminf}
\phi_u (r) \stackreb{r\to\infty}{=}
\sqrt{\xi}\left(1-c_{n,k}\frac{e^{-m_{\gamma}r}}
{\sqrt{m_{\gamma}r}}
+\cdots \right)
\\
\phi_d (r) \stackreb{r\to\infty}{=} \sqrt{\omega\xi}
\left(1+c_{n,k}\frac{2}{3(\Omega-\frac43\omega)}
\frac{e^{-m_{\gamma}r}}{\sqrt{m_{\gamma}r}}
+\cdots \right)
\\
\tilde{F}_1 (r) \stackreb{r\to\infty}{=}
\varepsilon_{n,k} c_{n,k}\frac{2m_{\gamma}^2}{\Omega}
\frac{e^{-m_{\gamma}r}}
{\sqrt{m_{\gamma}r}}+\cdots
\\
\tilde{F}_d (r) \stackreb{r\to\infty}{=} -\varepsilon_{n,k} c_{n,k}
\frac{2m_{\gamma}^2}{\sqrt{3}(\Omega -\frac43\omega)}\;
\frac{e^{-m_{\gamma}r}}
{\sqrt{m_{\gamma}r}}+\cdots
\ee
where $c_{n,k}$ will be determined below, while sign factors
$\varepsilon_{n,k}$ are
given by (\ref{sign}). In (\ref{nminf}) we
use the singular gauge and expressions
for dual field strengths $\tilde{F}_1$, $\tilde{F}_d$ related to
$f_1$, $f_d$ via
\be
\label{Ff}
\tilde{F}=-\frac{\varepsilon_{n,k}}{n_e r}
\frac{\partial f(r)}{\partial r}
\ee
where $n_e=\frac12$ and $n_e=\frac1{\sqrt{3}}$ for $f_1$ and $f_d$
respectively.

Clearly the leading asymptotic behavior in (\ref{nminf}) is determined by
the lightest gauge and scalar fields so the photon mass in (\ref{nminf})
coincides with smaller eigenvalue in formulas (\ref{r2mass}), (\ref{lam}),
i.e. $m_{\gamma}=(m_{\gamma})_{-}$ and $\Omega=\Omega_{-}$. In fact
(\ref{nminf}) already determines the relation between the fields
$\varphi_u$, $\varphi_d$, $\tilde{F}_1$ and $\tilde{F}_d$
and eigenvectors of the mass matrix (\ref{r2phmat})
$\varphi^{(-)}$, $\varphi^{(+)}$, $\tilde{F}^{(-)}$ and $\tilde{F}^{(+)}$
\be
\label{evs}
\delta\varphi_u= \cos{\beta_\varphi}\delta\varphi^{(-)} +
\sin{\beta_\varphi}\delta\varphi^{(+)}
\\
\delta\varphi_d= -\sin{\beta_\varphi}\delta\varphi^{(-)}+
\cos{\beta_\varphi}\delta\varphi^{(+)}
\ee
and
\be
\label{evg}
\tilde{F}_1=\cos{\beta_F}\tilde{F}^{(-)}+
\sin{\beta_F}\tilde{F}^{(+)}
\\
\tilde{F}_d=-\sin{\beta_F}\tilde{F}^{(-)}+
\cos{\beta_F}\tilde{F}^{(+)}
\ee
since $\beta_\varphi$ and $\beta_F$ are determined by ratios of
the coefficients
in front of leading asymptotics in (\ref{nminf})
\be
\label{beta}
\tan{\beta_\varphi}=\frac23\frac{\sqrt{\omega}}{\Omega -\frac43\omega}
\\
\tan{\beta_F}=\frac1{\sqrt{3}}\frac{\Omega}{\Omega -\frac43\omega}
\ee
Of course, the same relations
can be derived directly by diagonalization of the corresponding
mass matrices, see sect.~\ref{ss:r2spec}.
Using these formulas and asymptotic behavior in (\ref{nminf})
for the lightest scalar and vector fields at large $r$, one gets
\be
\label{linf}
\delta\varphi^{(-)} (r) = -c_{n,k}\frac{\sqrt{\xi}}{\cos{\beta_\varphi}}
\frac{e^{-m_{\gamma}r}}{\sqrt{m_{\gamma}r}}
+\cdots
\\
\tilde{F}^{(-)}(r) = \varepsilon_{n,k}c_{n,k}
\frac{2m_{\gamma}^2}{\Omega\cos{\beta_F}}
\frac{e^{-m_{\gamma}r}}{\sqrt{m_{\gamma}r}}+\cdots
\ee
It is clear that the effective vertex for $(n,k)$-string
would effectively depend in the leading order only upon the
lightest scalar and gauge
fields $\varphi^{(-)}$ and $A_\mu^{(-)}$ since only these fields
determine the large $r$ behavior of string interaction.
Following the same steps leading to effective
vertex for the ANO string in sect.~\ref{ss:intano} and
using again the propagators (\ref{sprop}), (\ref{gprop}) we arrive
to the following expression
\be
V_{n,k} \sim \int DX(\bsigma)\exp \left\{-\int d^2\sigma 2\sqrt{2\pi}
c_{n,k} \left[
\frac{\sqrt{\xi}}{\cos{\beta_\varphi}}
\left(\delta\bar{\varphi}^{(-)}+\delta\varphi^{(-)}\right)
+\frac{\varepsilon_{n,k}}{g^2\Omega\cos{\beta_F}}
F_{\mu\nu}^{(-)}n_{\mu\nu} + \dots
\right]\right\}
\label{nmV}
\ee
up to normalization factor. Let us finally fix the constants $c_{n,k}$ here.
To do this rewrite expression (\ref{nmten}) for the string tension
in the form
\be
\label{tdf}
T_{n,k}(\varphi) = 2\pi \left|n |\varphi_u|^2 +k|\varphi_d|^2\right|
\ee
Expanding in (\ref{tdf}) the
scalar fields around their VEV's one can extract the linear term
in $\delta\varphi^{(-)}$ using relations (\ref{evs}). Comparing it with
the expression linear in $\delta\varphi^{(-)}$ in the exponent
in (\ref{nmV}) we find for the coefficients $c_{n,k}$
\be
\label{anm}
c_{n,k}=\varepsilon_{n,k}\sqrt{\frac{\pi}{2}}\cos^2{\beta_\varphi}\left(n-
k\tan{\beta_\varphi}\sqrt{\omega}\right)
\ee
Substituting this back to (\ref{nmV}) and taking into account (\ref{tdf})
we finally arrive to the following effective $(n,k)$-string vertex
\be
\label{nmVf}
V_{n,k} =  \int DX(\bsigma)\exp \left\{-2\pi \int d^2\sigma\left[
 \sqrt{\det g_{\rm ind}} \left| n|\varphi_u|^2 +k|\varphi_d|^2\right|
\right.\right.
\\
\left. +\frac{\cos^2{\beta_\varphi}}{ g^2\Omega\cos{\beta_F}}
\left.\left(n-k\tan{\beta_\varphi}\sqrt{\omega}\right)
F_{\mu\nu}^{(-)}n_{\mu\nu} + \dots
\right]\right\} ,
\ee
where, as in (\ref{string}) we have already taken into account motion of
string. Here only $F^{(-)}_{\mu\nu}$ enter the exponential because
only this component survives at large distances. However, on general grounds
it is natural to
expect that the effective vertex depend on the following combination
of two gauge fields: ${\bf q}_{n,k}{\bf F}_{\mu\nu}$, where
we defined vector field strength as
\be
\label{vecF}
{\bf F}_{\mu\nu}={\bf e}_1 F_{1\mu\nu}+{\bf d}\,\sqrt{3}\,F^{(d)}_{\mu\nu}.
\ee
Taking this into account we can conjecture that the
effective vertex has the form
\be
V_{n,k}
= \int DX(\bsigma)\exp \left\{-2\pi \int d^2\sigma\left[
 \sqrt{\det g_{\rm ind}} \left| n|\varphi_u(X(\bsigma))|^2 +
k|\varphi_d(X(\bsigma))|^2\right|
\right.\right.
\\
\left. + \frac1{g^2\Omega}
\left.\ {\cos^2\beta_\varphi\over\cos^2\beta_F}\
{\bf q}_{n,k}\cdot
{\bf F}_{\mu\nu}(X(\bsigma))n_{\mu\nu}
\right]\right\}
\label{nkV}
\ee
Exponential  here depends only on the component of the
field strength, directed along the charge
${\bf q}_{n,k}$ of the $(n,k)$-string. The extra piece that
appears in (\ref{nkV}) as compared to (\ref{nmVf}) associated
with $F^{(+)}_{\mu\nu}$ component of the gauge field and do not
contribute at large distances. We are not going to use
the effective vertex in the (conjected) form (\ref{nkV}) below  and
write it down only in the sake of completeness.

Let us finally compute the interaction potential of strings at large distances,
technically for this purpose it is easier to use effective vertex
in the form (\ref{nmV}).
Expanding  $\exp\left(-\sum_{n,k}V_{n,k}\right)$ in powers of $V_{n,k}$
we keep again only the term $\langle
V_{n_1,k_1}V_{n_2,k_2}\rangle$.   Calculation of this
correlation function to the leading order in coupling constant using
tree level propagators (\ref{sprop}), (\ref{gprop})
leads to the following interaction potential of $(n,k)$-strings
\be
\label{nmU}
U_{1,2}=-\frac{16}{3}\sqrt{2\pi}\xi c_1 c_2
\frac{(\Omega-\omega)^2+\frac13\omega^2}{\Omega(\Omega-
\frac43\omega)^2}\frac{e^{-m_{\gamma}R}}{\sqrt{m_{\gamma}R}}
\left[1-\varepsilon_1 \varepsilon_2 \right]
\ee
where $c_1\equiv c_{(n_1,k_1)}$ for $(n_1,k_1)$-string and
$c_2\equiv c_{(n_2,k_2)}$ for $(n_2,k_2)$-string are given by (\ref{anm}).
Quite similar to the case of ANO strings the first term in square
brackets here comes from the exchange by two scalars $\varphi_u$ and
$\varphi_d$ and this interaction always gives rise to attractive potential.
The second term arises due to exchange by two photons $A_{1\mu}$ and
$A_{\mu}^{(d)}$, the sign of this contribution is determined by product
of the sign factors (\ref{sign}) for two strings.
In particular, for $(n_1+\omega k_1)(n_2+\omega k_2)>0$ the photon
exchange gives repulsion which cancels the attraction produced by
the scalar exchange and in this case strings do not interact.
If instead $(n_1+\omega k_1)(n_2+\omega k_2)<0$ the interaction potential
is attractive so that two strings form a stable bound state.

\subsection{Lattice of (n,k)-strings
\label{ss:lattice}}

Now let us use the interaction potential (\ref{nmU}) to verify
the existence and stability of the $(n,k)$-string
solutions. We are going to determine which strings among the solutions
to the BPS first order equations (\ref{1foe}) are stable and which
do not exist as BPS states.
 The similar approach was used in \cite{RSVV,RV}
to study BPS soliton states in two dimensions and various dyon states in
Seiberg-Witten theory.

Suppose first, one has two BPS strings $(n_1,k_1)$ and $(n_2,k_2)$ with
$(n_1+\omega k_1)(n_2+\omega k_2)>0$. Then according to
(\ref{nmU}) these strings do not interact, it means that the
bound state of these strings $(n_1+n_2,k_1+k_2)$ exists as a BPS solution to
the first order equations (\ref{1foe}) but it is only {\em marginally} stable.
The tension of this string is given according to (\ref{nmten}) by
sum of the tensions of its "components"
\be
\label{sumten}
T_{1\oplus 2}=T_{1}+T_{2}
\ee
where $T_s\equiv T_{(n_s,k_s)}$
\footnote{Note that we restrict ourselves
only to real and positive values of parameter $\omega$ (\ref{x}).
For complex $\omega$ one would have
far more complicated lattice of $(n,k)$-strings. In particular the mass
formula (\ref{nmten}) suggests that we probably
can have stable BPS bound states
at complex $\omega$ which then decay at real $\omega$. Thus the surface
$\Im\, \omega=0$ should be
the curve of marginal stability (CMS) for these bound
states.}.
If instead $(n_1+\omega k_1)(n_2+\omega k_2)<0$, two components attract
each other and form a stable bound state. However we do not know {\em a
priori} if this bound state is BPS saturated or not, all we know is that
its tension is within the bounds
\be
\label{nBPSten}
|T_1-T_2|<T_{1\oplus 2}<T_1+T_2
\ee
where the lower bound is given by the BPS formula (\ref{nmten}) while the
upper bound follows from the potential (\ref{nmU}).

From (\ref{sign}) we see that interaction of strings depends
on the value of parameter $\omega$. Therefore
let us first distinguish physically different regions for this
parameter. The masses of W-bosons are classically given by (\ref{wbps})
and, as follows from (\ref{Higgs}), (\ref{r2phi}) and (\ref{x}),
they are proportional to
\be
\label{wmassz}
\left| Mz \right|,\;\left| \frac{M}{z} \right|,\;
\left| M\left(z-\frac1z\right)\right|
\ee
where instead of $m_1$ and $m_2$ we have introduced
new variables $M$ and $z$ defined by
\be
M^2=(2m_1+m_2)(2m_2+m_1)
\\
z^2=\omega
\label{z}
\ee
One can see from (\ref{wmassz}) that at special values
$\omega=0$, $\omega=1$ and $\omega=\infty$ one of the
W-boson masses vanishes and it corresponds to the restoration of
corresponding $SU(2)$ gauge subgroup. However, as we mentioned before
this is correct only classically and in quantum theory the
$SU(2)$ subgroups are never restored for the theories with
$N_f<4$ (in the next section we consider the theories with four and
five flavors
where the $SU(2)$ subgroup is restored at $\omega=1$). Instead
the $SU(2)$-subsector runs into strong coupling and W-bosons never become
massless \cite{SW1,SW2}.
Hence, it is clear that three regions around values $\omega=0,1$ and $\infty$
are in fact strongly coupled and we cannot use our semiclassical analysis
there, therefore one has two separated weak coupling regions
\be
\label{omreg}
0<\omega <1 \;\;\; {\rm and}\;\;\; 1<\omega <\infty
\ee
and we cannot pass from one region to another within weak coupling regime
keeping real $\omega$. However, in the effective low-energy Abelian theory
these three points look differently: at $\omega=0$ and $\omega=\infty$ one
of the photons becomes massless (see (\ref{lam})), which is not true for
$\omega=1$. In fact, in theory with classically restored $SU(2)$ subgroup
value $\omega=0$  ($\omega=\infty$) corresponds to the Argyres-Douglas
point \cite{AD,APSW,GVY,Y01} where the monopole/dyon vacuum (${\tt r}=1$ vacuum)
collides with
charge vacuum (${\tt r}=2$). Thus the values $\omega=0$ and $\omega=\infty$
correspond to strong coupling even in the Abelian low energy theory.
On the contrary, nothing special happens in the Abelian theory
(\ref{qed}) at $\omega=1$, the low energy theory does not "feel" the
restoration of $SU(2)$ subgroup. This means that in the
low energy description we have only one region $0<\omega <\infty$
instead of two regions (\ref{omreg}).

\begin{figure}[tb]
\epsfysize=9cm
\centerline{\epsfbox{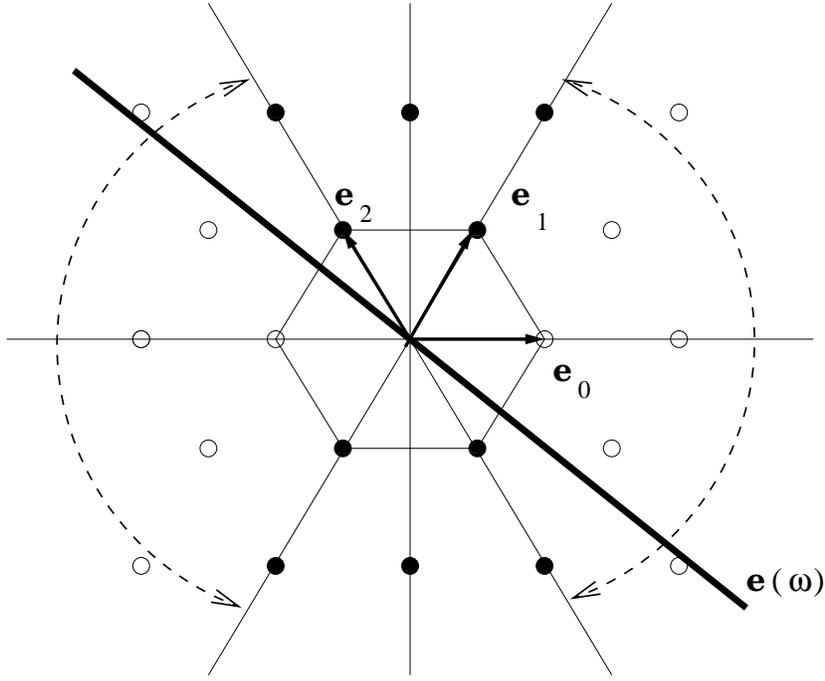}}
\caption{\sl Lattice of string solutions for $0<\omega<1$. The non-BPS strings
are drawn by white dots while the BPS strings are depicted by black dots.
All BPS
strings are marginally stable bound states of "fundamental" ${\bf e}_1$- and
${\bf e}_2$-strings, corresponding to two "closest" roots of
$SU(3)$ algebra. The non-BPS strings can be "crossed" by straight line
$n+\omega k =0$ or ${\bf e}(\omega)$ when
one moves parameter $\omega$ in the region $0<\omega<\infty$.}
\label{fi:latw}
\end{figure}

Consider the straight line
$n+\omega k =0$ as is shown on the lattice of $(n,k)$-strings on
fig.~\ref{fi:latw}.
This line is directed along the vector
\be
\label{zeroline}
{\bf e}(\omega)=\omega{\bf e}_1-{\bf e}_2
\ee
orthogonal to the vector $\bphi$, see (\ref{2phi}).
If charges of two BPS strings are both on the same side out of the line
$n+\omega k=0$ then, according to (\ref{nmU}) and (\ref{sign}), these
strings do
not interact and form marginally stable state.
If instead they are from the opposite sides of this line, they
attract each other and form a stable bound state.

\begin{figure}[tb]
\epsfysize=10cm
\centerline{\epsfbox{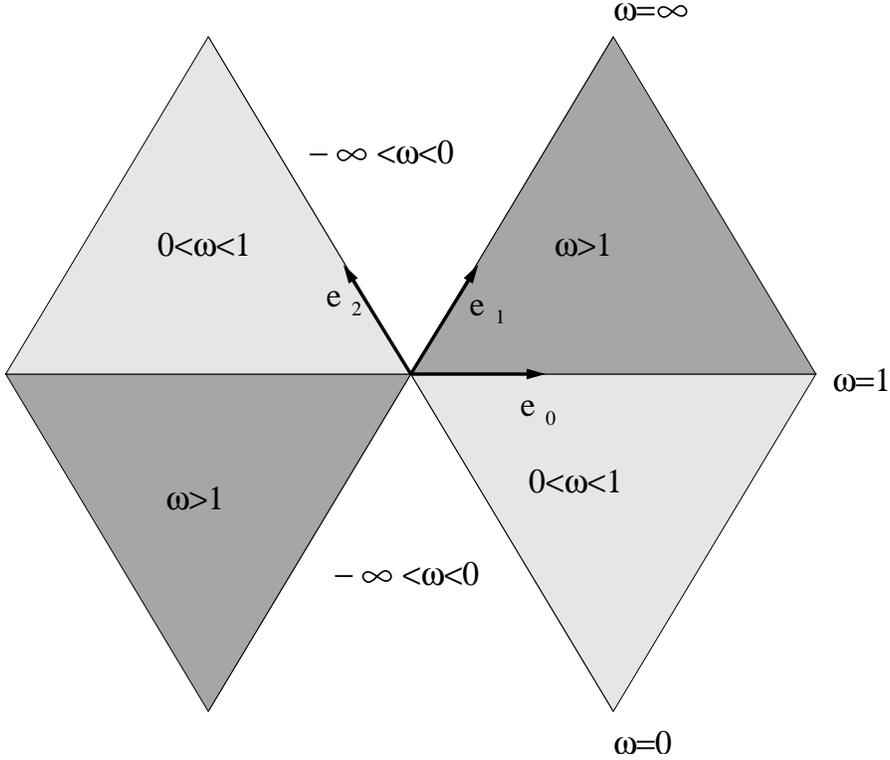}}
\caption{\sl Different values of the parameter $\omega$. Values $0<\omega<1$ are
restricted by straight lines $\omega=0$ along the vector ${\bf e}_2$ and $\omega=1$
along ${\bf e}_0$, values $\omega>1$ are between the lines $\omega=1$ along ${\bf e}_0$
and $\omega=\infty$, which is along the vector ${\bf e}_1$.}
\label{fi:omegas}
\end{figure}

At $\omega=0$ the straight line ${\bf e}(\omega)$ is directed along
${\bf e}_2$ and as we increase $\omega$ moves anti-clockwise
reaching the vector ${\bf e}_0$ at $\omega=1$, see
fig.~\ref{fi:latw}. Moving it further, one reaches the vector
${\bf e}_1$ at $\omega=\infty$.
As the line directed along the vector
${\bf e}(\omega)$ hits any knot on the root lattice the BPS
formula (\ref{nmten})
\be
\label{nmt} T^{BPS}_{n,k}=\,2\pi \xi
|n+\omega k|
\ee
gives zero for the BPS bound of corresponding
string.  Thus if within the two $2\pi/3$ angles between the
vectors $-{\bf e}_2$ and ${\bf e}_1$ (${\bf e}_2$
and $-{\bf e}_1$) on fig.~\ref{fi:latw} (see also "shadowed" regions
at fig.~\ref{fi:omegas},
where this sector is shown explicitly) all stings were BPS saturated
one would get {\em infinitely many} strings becoming tensionless.
To avoid this we have to accept that all these strings
are actually {\em non-BPS} states at real positive $\omega$ (cf. \cite{RV}).
This is in accordance with our conclusion that BPS solutions exist only
if  $n$ and $k$ are both positive or negative, see (\ref{fir0}).

Thus, one gets the following picture for the $(n,k)$-strings:
${\bf e}_1$- and ${\bf e}_2$-strings ($-{\bf e}_1$
and $-{\bf e}_2$), with charges proportional to the two closest roots
on the $SU(3)$ root lattice (see
fig.~\ref{fi:su3}) are the lightest BPS stable "elementary"
strings. Both strings are from the same side of
the line $n+\omega k=0$, see
fig.~\ref{fi:latw}. Therefore all strings within two
$\frac{\pi}{3}$ angles between vectors ${\bf e}_1$ and ${\bf
e}_2$ ($-{\bf e}_1$ and $-{\bf e}_2$) labeled by black circles at
fig.~\ref{fi:latw} exist as a BPS solutions of (\ref{1foe}) but
they are marginally unstable.  Instead all strings labeled at
fig.~\ref{fi:latw}, by white circles can be considered as bound
states of $q{\bf e}_1$ and $-p{\bf e}_2$ strings (where $q$, $p$
are integers and $qp>0$),
which are from {\em different} sides of the line $n+\omega k=0$. In this case
the components are in attractive channel, hence, the composite strings
are stable, but as we showed before they are {\em not} BPS saturated.

Note as an additional check that $(1,1)$ string is BPS marginally
stable state.
This is in complete agreement with our discussion at the end of
sect.~\ref{ss:r2strings} where we have shown that this string should
exist as a BPS solution of (\ref{1foe}) in some region around
$\omega=1$.
Note also that ${\bf e}_1$ and ${\bf e}_2$ strings in fact do
not become tensionless
at $\omega=0$ and $\omega=\infty$ respectively as suggested by BPS formula
(\ref{nmt}).
As we already mentioned "boundary" values $\omega=0$ and $\omega=\infty$
corresponds to strong coupling where one of the photons becomes massless
and classical BPS mass formulas should be modified.


Note also that there is a hidden symmetry of a W-boson spectrum
(\ref{wmassz})
\footnote{This symmetry is very similar to a
symmetry in Toda chain integrable system, which has already
appeared in the context of SUSY gauge theories as an elegant way
to formulate the Seiberg-Witten exact solution \cite{GKMMM}.}
\be
M = {\rm fixed}
\\
\label{zsym}
z\to\frac1z
\ee
and let us check now that the spectrum of light particles in ${\tt r}=2$ vacuum
also respects this symmetry. The eigenvalues of mass matrix are given by
eqs. (\ref{r2mass}), (\ref{lam}), and in terms of variables (\ref{z}) they
get the form
\be
\label{phmz}
(m^2_{\gamma})_{\pm}=\frac23 g^2\mu M\left(\frac1z+z\pm
\sqrt{\frac1{z^2}-1+z^2}\right)
\ee
Therefore we conclude that the low energy spectrum is also invariant under the
symmetry (\ref{zsym}).
Now consider the spectrum of BPS $(n,k)$ strings given by (\ref{nmt}),
which in terms of $M$ and $z$ (\ref{z}) reads
\be
\label{tenz}
T^{BPS}_{n,k}=4\pi\mu M\left|\frac{n}{z}+ kz\right|
\ee
We see that it is indeed invariant under transformation (\ref{zsym})
together with exchange in the string quantum numbers
\be
(n,k)\leftrightarrow (k,n)
\label{nmz}
\ee

To conclude this section let us sum up the picture of the
monopole confinement at ${\tt r}=2$ vacuum.
The ${\bf e}_1$-monopole-antimonopole pair
is confined by the ${\bf e}_1$-string, while
${\bf e}_2$-monopole-antimonopole pair is confined by the
${\bf e}_2$-string. The ${\bf e}_0$-monopole-antimonopole pair is confined by
the ${\bf e}_0$-string which is stable bound state of
${\bf e}_1$- and $-{\bf e}_2$-strings. We see that one gets three
monopole-antimonopole meson states instead of one. As we have
already explained, this "multiplicity" reflects the {\em Abelian}
nature of confinement in Seiberg-Witten theory. For generic values of
$\omega$ the masses of all three  mesons are different (they are
determined by different values of string tension).

Finally, let us note, that we predict presence of the non-BPS
stable strings which can be considered as bound states
of "elementary" BPS strings in attractive channel.
These strings can form an "exotic" multi-monopole-multi-antimonopole
mesons. The example of such string is
(-1,2) string which is a stable non-BPS bound
state of $(-{\bf e}_0)$- and ${\bf e}_2$-string.
This string bound
${\bf e}_2$-monopole and
${\bf e}_0$-antimonopole with
${\bf e}_2$-antimonopole and
${\bf e}_0$-monopole to form an "exotic" meson.
Presence of these "exotic" mesons also reflects the
Abelian nature of confinement in Seiberg-Witten theory
\cite{St,VY}. However, we do not expect such states to appear
in theory with non-Abelian confinement.

\section{$SU(2)$ gauge symmetry restoration in the theory with
four and five flavors
\label{ss:nonabelian}}

Finally, let us turn to the most physically interesting examples --
the $SU(3)$ gauge theory with $N_f=4$ and especially $N_f=5$ flavors.
As we already explained, the motivation to study
such theories is that at certain submanifolds in their moduli space
non-Abelian $SU(2)$ gauge symmetry is restored
and one can study what happens to confinement and flux tubes
in this regime. The reason for restoration of non-Abelian $SU(2)$ gauge
symmetry in quantum theory is that at $N_f=4$ and $N_f=5$ the
corresponding $SU(2)$ subsectors are {\em not} asymptotically free
and they remain to be non-Abelian in weak coupling
regime at $m_A\gg\Lambda$ (cf. \cite{APS}).

Let us take a closer look at ${\tt r}=2$ vacua of these theories. The theory
with $N_f=4$  has six ${\tt r}=2$ vacua while
the theory with $N_f=5$ has ten ${\tt r}=2$ vacua, see (\ref{nrvac}).
Now consider the point in the parameter space where masses of
all flavors $m_A$ become equal. At this point six (ten) vacua of the
theory with $N_f=4$ ($5$) flavors collide with the two-fold
effect.  First, as we already discussed in sect.~\ref{ss:colide},
the Higgs branch of dimension $8(N_f-2)$ develops. It touches the
Coulomb branch at the point \be \label{root} a_3=0,\;
a_8=-\sqrt{6}m, \ee where $m=m_A$ is now common value of the
quark masses, see (\ref{r2vev}).

Second, since $a_3=0$ the $SU(2)$ gauge subgroup acting in the
upper left corner of the adjoint field matrices is classically
restored, see (\ref{phigen})
\footnote{Strictly
speaking this is true at energies much larger
then quark VEV's which break gauge group completely
at the scale of order of $\sqrt{\mu m}$, see discussion below.}.
this restoration is also preserved in  quantum theory.
Namely, the $SU(2)$-subsector for $N_f=4$ is conformal and the
coupling constant does not run, being of the order of
\be
\label{4coup}
g^2\sim \frac1{\log{(m/\Lambda)}}\ll 1
\ee
(with $\Lambda\equiv\Lambda_{SU(3)}$; the $SU(3)$ theory with $N_f=4$ is not
conformal), frozen at the scale $m$ where $SU(3)$ group is broken down to
$SU(2)\times U(1)$.

For $N_f=5$ theory the $SU(2)$-subsector has "zero charge" in
the infrared limit. The coupling constant at large distances
is of the order of
\be
\label{5coup}
g^2\sim -\frac1{\log{(\Delta m\Lambda/m^2)}}\ll 1,
\ee
frozen at  lower scale of quark mass difference $\Delta m$
(note that $\Lambda_{SU(2)}=m^2/\Lambda_{SU(3)}\equiv m^2/\Lambda $
for $N_f=5$).

We consider here special submanifold of the Higgs branch
which admits the BPS flux tubes (cf. \cite{HSZ,GS}). We call this
submanifold  origin of the Higgs branch. One point on
this submanifold  corresponds to non-zero $u$-component of the
first flavor and non-zero $d$-component of the second flavor
given by
\be
\label{orhi} \langle\tilde{u}_1u^1\rangle =
\langle\tilde{d}_2d^2\rangle
=3\mu m =\frac{\xi}{2}
\ee
while all other components are zero. This
is a continuation of ${\tt r}=2$ vacuum considered in previous
sections to the case of
equal quark masses, see (\ref{vacr2}).
Other points in the origin of Higgs branch are given by $SU(N_f)$
flavor rotation of (\ref{orhi}). The dimension of the origin of the
Higgs branch is $4(N_f-2)$. To see this note that VEV's of $u$ and
$d$-quarks in ${\tt r}=2$ vacuum break $SU(N_f)$ symmetry down to
$SU(N_f-2)$. Thus the number of "broken" generators is
$\dim SU(N_f) - \dim SU(N_f-2) = 4(N_f-1)$
and four phases are ``eaten'' by Higgs mechanism.

Other points on the $8(N_f-2)$ dimensional Higgs
branch correspond to non-zero VEV's of massless moduli fields, and
these points do not admit BPS strings. In particular, the ANO
strings on Higgs branch were studied in \cite{Y99}, they
correspond to limiting case of type I strings with the
logarithmically thick tails associated with massless scalar
fields. Moreover, the infinitely long strings
do not exist \cite{RTT,Y99} and we do not discuss here strings
in generic points on Higgs branch.

Now let us focus on the strings arising in some particular vacuum
at the origin of Higgs branch. By flavor rotation one can
always put this vacuum to the form (\ref{orhi}). To study the
strings at this vacuum we can apply results of the previous
section and take the limit $\omega\to 1$.

Suppose we approach the value $\omega=1$.
%
Let us make a closer look at what happens to the lattice of strings
from fig.~\ref{fi:lattice}.
It is easy to see that ${\bf e}_0$-string corresponds to winding around
the $U(1)$ factor associated with the generator $\lambda_3$, see
(\ref{lam3}). This generator belongs to the restored $SU(2)$
subgroup. However, $SU(2)$ group does not admit flux tubes
just because $\pi_1 (SU(2))=0$. Now  it is clear that ${\bf e}_0$-string
becomes unstable at $\omega\to 1$. In fact, the winding just shrinks to
zero once the group manifold becomes $3$-sphere instead of a
circumference.

What happens physically is that ${\bf e}_0$-string is broken by
${\bf e}_0$-monopole-antimonopole production which does not
cost any energy in the limit of equal quark masses. To see  this
note that mass of the ${\bf e}_0$-monopole vanishes in this limit.
Namely, the BPS ${\bf e}_0$-monopole mass in
weak coupling regime has the form \cite{SW1}
\be
\label{e0mm}
m^M_{e_0}=\sqrt{2}\left|\frac{a_3}{g^2}\right|
\ee
and vanishes at $a_3\to 0$.

Thus, we see that  ${\bf e}_0$-string becomes unstable
and disappears from the spectrum.
Let us discuss now what happens to confined states
when we tend to zero the differences of quark masses $\Delta m_{AB}=
m_{A}-m_{B}$
(assuming that they are all
of the same order $\Delta m_{AB} \sim \Delta m $). As we already
explained at $\sqrt{\mu m}\ll \Delta m \ll m$ one has weak
coupling and the mass of ${\bf e}_0$ W-boson is of the order of $\Delta m$,
while the mass of ${\bf e}_0$-monopole is given by (\ref{e0mm}).
When we put $\Delta m$ well below  $\sqrt{\mu m}$ the
mass of ${\bf e}_0$ W-boson (together with the masses of photons)
is determined by complete breaking of $SU(2)$ gauge subgroup
by the quark VEV's (instead of adjoint VEV $\langle a_3\rangle\sim\Delta m$), and
it is frozen at the value of order of $\sqrt{\mu m}$. As we reduce
$\Delta m$ the ${\bf e}_0$-monopole becomes {\em lighter} than
${\bf e}_0$ W-boson. This shows that eqs.(\ref{4coup}), (\ref{5coup})
are no longer valid and we enter into strong coupling regime. To
understand this, note that presence of simultaneously light
quarks (which become massless quark moduli at $\Delta m=0$) and
light monopole means approaching the point of Argyres-Douglas
type \cite{AD,APSW,GVY,Y01}. Apparently as ${\bf e}_0$-monopole becomes
lighter and lighter the ${\bf e}_0$-string becomes more and more
unstable.

As we already discussed, presence of $N_c-1$
different strings associated with each $U(1)$ factor of
$SU(N_c)$ gauge group broken down to $U(1)^{N_c-1}$ leads
to unwanted multiplicity in the hadron spectrum of \N2 QCD
which we do not expect in ordinary QCD \cite{DS}, and this reflects
the $U(1)$ nature of the confinement in Seiberg-Witten theory.
In particular, one has generically two "elementary"
strings in $SU(3)$ gauge theory, namely ${\bf e}_1$-string and
${\bf e}_2$-string in ${\tt r}=2$ vacuum. As we already explained
in the last section this leads to the presence of three
monopole-antimonopole meson states formed by
${\bf e}_1$, ${\bf e}_2$ and ${\bf e}_0$ strings.

However, in the theory with  $N_f=4$ and $N_f=5$ flavors with
equal masses with the non-Abelian $SU(2)$
subgroup of gauge symmetry  restored
at ${\tt r}=2$ vacuum ${\bf e}_0$-string becomes unstable.
This has two-fold effect on the meson spectrum. First
${\bf e}_0$-monopole-antimonopole meson formed
by ${\bf e}_0$-string with ${\bf e}_0$-monopole and antimonopole
attached to its ends acquire large width.
Eventually at $\Delta m\to 0$ it becomes unobservable as a
resonance state and disappear
from the spectrum.

Second, ${\bf e}_1$ and ${\bf e}_2$ monopole-antimonopole mesons formed
by ${\bf e}_1$ and ${\bf e}_2$ strings respectively become mixed with the
transition amplitude of order of one because ${\bf e}_1-{\bf e}_2=
{\bf e}_0$. Hence, these two mesons form two splited states, more massive
state acquires large width and also disappears from the spectrum.
As a result we end up with only one set of monopole-antimonopole
meson Regge trajectories.

Hence, we see that there is
no longer unwanted multiplicity in the hadron spectrum.
Although confinement in this theory is
due to the presence of Abelian ${\bf e}_1$-string
($\sim {\bf e}_2$-string), the hadron spectrum
has multiplicity which we expect in a theory with non-Abelian
confinement.

Note also that extra multi-monopole-muti-antimonopole "exotic"
states formed by stable non-BPS strings (see sect. 5.3)
also disappear in the equal mass limit in  theories
with $N_f=4$ or $N_f=5$.

To conclude this section let us  make a comment on so called
semilocal strings (see \cite{AV} for a review). These are
string-like solutions which interpolate between BPS strings and
two-dimensional sigma model instantons lifted to four dimensions. Instead of
ordinary BPS strings these solutions have power fall-off at large
$r$. The semilocal strings arise in the models with additional global
symmetry. In the theories with $N_f=4$ and $N_f=5$ considered in this
section one has broken
global $SU(N_f)$ symmetry at the origin of Higgs
branch which leads to presence of massless Goldstone scalars
responsible for possible power behavior of string  solution at
large $r$.  The presence of semilocal string manifests itself as a
zero mode of the ordinary BPS string.

Therefore it would be interesting to investigate how ordinary BPS strings
with exponential fall-off at infinity studied in this paper
turn into semilocal strings as we aproach the point of $SU(2)$
gauge symmetry restoration. One comment here is that
 as it is mentioned
in sect.~\ref{ss:anor1} our strings are BPS saturated only in the leading order
in $\mu/m$ when perturbation of superpotential (\ref{supot})
reduces to the FI term and \N2 supersymmetry is not broken. Already in
the next to leading order in $\mu/m$ these strings
turn into type I strings \cite{VY}. For type I strings
the zero mode associated with the transverse size of the
 string (which is
similar to the instanton size)
becomes a positive mode. It means that
the transverse size of the semilocal string although
grows as compared to the size of the ordinary string
($\sim 1/\sqrt{\mu m}$) is still bounded by $1/\mu^{3/4}m^{1/4}$,
and the semilocal string does not grow infinitely thick.

\section{Conclusion}

In this paper we have studied the flux tubes in softly broken \N2 QCD
arising in quark vacua of $SU(N_c)$ gauge theory with $N_f$ flavors at
weak coupling, in particular focusing on the theory  with
$SU(3)$ gauge group. These magnetic flux tubes are responsible
for the confinement of monopoles. Of course, it would be more
desirable to study confinement of quarks arising in dyon
vacua at strong coupling via electric flux tubes. Still
we believe that the two problems are directly related due to
monodromies \cite{SW1,SW2} and it is useful to start with
simpler problem of monopole confinement in quark vacua, which one
can keep at weak coupling.

Generically $SU(3)$ gauge group is broken down to $U(1)^2$, thus
in each \1N vacuum two "elementary" flux tubes arise. In ${\tt r}=1$
vacua (when only $u$-quark condense) these are magnetic ${\bf u}$-string
and electric ${\bf e}_2$-string. This hybrid phase arises since
classically the gauge group is broken down only to $SU(2)\times U(1)$.
However on quantum level the $SU(2)$ factor is further broken
down to another $U(1)$ due to the Seiberg-Witten strong coupling mechanism. This
results in the condensation of ${\bf e}_2$-monopole (dyon) and
formation of electric ${\bf e}_2$-string. We have shown that
strings in ${\tt r}=1$ vacua are the standard BPS ANO vortices
in the limit of small adjoint masses $\mu$.

In ${\tt r}=2$ vacua with non-zero VEV's of $u$- and $d$-quarks two
"elementary" BPS strings are magnetic ${\bf e}_0$ and ${\bf e}_2$-string
(at the values $0<\omega <1$ of VEV's ratio
$\omega = \langle\tilde{d}_2d^2\rangle/
\langle\tilde{u}_1u^1\rangle$).
These strings are generalization of standard ANO vortices, and
they involve two gauge potentials interacting with two scalar
fields satisfying BPS first order eqs.~(\ref{1foe}). We have demonstrated
that bound states of multiple ${\bf e}_1$-string and multiple
${\bf e}_2$-string are marginally stable strings while
bound states of multiple ${\bf e}_1$-string and multiple
${\bf e}_2$-antistring are stable non-BPS strings.

Finally we have considered the theory with $N_f=4$ and $N_f=5$ in the limit
of equal quark masses. In this limit $SU(2)$ subgroup
of the gauge group is not broken even on quantum level in ${\tt r}=2$
vacuum. We have shown that in  the limit of equal quark masses
${\bf e}_0$-string becomes unstable and disappears from the spectrum.
The monopole confinement is due to remaining ${\bf e}_1$-string which
is not distinguishable in this limit from ${\bf e}_2$-string.  This
mechanism eliminates the unwanted multiplicity of the hadron
spectrum  generically present in Seiberg-Witten theory. The
meson spectrum looks as expected in a theory with {\em non-Abelian}
confinement. Namely, one gets the only set of
${\bf e}_1$-monopole-antimonopole meson Regge trajectories
(or, equivalently, ${\bf e}_2$-monopole-antimonopole meson).

Something peculiar happen to "baryons" in this limit.
Of course we cannot really talk about baryons formed by monopoles
since monopoles have zero baryon number. Still we can
consider neutral baryon-like states formed by ${\bf e}_0$, ${\bf e}_2$ and
$\bar{\bf e}_1 = - {\bf e}_1$ monopoles connected by
 $-{\bf e}_1$ and ${\bf e}_2$  strings in a chain-like
configuration.  Generically when $SU(3)$ gauge group is broken
down to $U(1)^2$ there are six different "baryon" states of this
kind. However, in the limit of equal quark masses when $SU(2)$
gauge subgroup is restored  and ${\bf e}_0$-string disappears
these "baryons" reduce to the only meson formed by
${\bf e}_1$-monopole and ${\bf e}_1$-antimonopole.
It is possible because
monopoles have zero baryon number.

This is not exactly what one expects in a theory with non-Abelian
confinement. In a theory with non-Abelian confinement
one expects to have one set of meson and one set of
baryon Regge trajectories. In the theory at hand in the limit of
equal quark masses "baryon" states disappear and we are left with
only one set of meson trajectories. It would be interesting to
study the mechanism of non-Abelian confinement via Abelian flux
tubes suggested in this paper for the case of monopole
condensation and quark confinement. In particular, this may
allow us to find what happens with baryon multiplicity upon
non-Abelian gauge subgroup restoration.

\subsection*{Acknowledgements}

We are grateful to P.~Arseev, N.~Fedorov, A.~Gorsky, K.~Konishi, A.~Morozov,
V.~Rubakov,  M.~Shifman and D.~Tong
for  useful discussions. Our special thanks are to A.~Vainshtein
for numerous illuminating discussions and critical comments on the earlier
version of this paper.
A.~Y. would like to thank the Theoretical
Physics Institute of the University of Minnesota
where part of this work has been done for hospitality and support.
The work was supported by Russian Foundation
for Basic Research under the grants No~00-02-16477 (A.M.) and
No~02-02-17115 (A.Y.), by INTAS grant No~00-00334 and by the
grants for support of scientific schools
No.~00-15-96566 (A.M.) and No.~00-15-96611 (A.Y.).

\section*{Appendix. SU(3) conventions and beta-functions}

The variables $\phi_i$ are associated with the basis vectors of the fundamental
representation ${\bf 3}$ (see fig.~\ref{fi:su3}):
$\phi_i = \bphi\bmu_i$, $i=1,2,3$, where $\bphi$ is
a VEV vector in Cartan plane. Their relation with Cartesian co-ordinates
$\bphi = (a_3,a_8)$ is given by
\be
\langle\Phi\rangle = \left(
\begin{array}{ccc}
  \phi_1 & 0 & 0 \\
  0 & \phi_2 & 0 \\
  0 & 0 & \phi_3
\end{array}\right) =
{a_3\over 2}
\left(
\begin{array}{ccc}
  1 & 0 & 0 \\
  0 & -1 & 0 \\
  0 & 0 & 0
\end{array}\right) +
{a_8\over 2\sqrt{3}}
\left(
\begin{array}{ccc}
  1 & 0 & 0 \\
  0 & 1 & 0 \\
  0 & 0 & -2
\end{array}\right) \equiv \lambda_3a_3 + \lambda_8a_8
\ee
i.e. $\langle\Phi\rangle$ is decomposed over well-known diagonal
Gell-Mann matrices
\be
\label{lam3}
\lambda_3 = \frac{1}{2}\left(
\begin{array}{ccc}
  1 & 0 & 0 \\
  0 & -1 & 0 \\
  0 & 0 & 0
\end{array}\right) = \2(\balpha_{12}-\balpha_{2}) = \2\balpha_{1}
= \2(\bmu_1-\bmu_2)
\ee
and
\be
\label{lam8}
\lambda_8 = \frac{1}{2\sqrt{3}}\left(
\begin{array}{ccc}
  1 & 0 & 0 \\
  0 & 1 & 0 \\
  0 & 0 & -2
\end{array}\right) = {1\over 2\sqrt{3}}(\balpha_{12}+\balpha_{2})
= {1\over 2\sqrt{3}}(\bmu_1+\bmu_2-2\bmu_3)
\ee
where we have used the following identification between the vectors
in Cartan plane (see fig.~\ref{fi:su3}) and diagonal matrices
\be
\label{e1m}
\balpha_{12} = \left(
\begin{array}{ccc}
  1 & 0 & 0 \\
  0 & 0 & 0 \\
  0 & 0 & -1
\end{array}\right) = \bmu_1-\bmu_3
\ee
and
\be
\label{e2m}
\balpha_{2} = \left(
\begin{array}{ccc}
  0 & 0 & 0 \\
  0 & 1 & 0 \\
  0 & 0 & -1
\end{array}\right) = \bmu_2-\bmu_3
\ee
(vectors (\ref{e1m}) and (\ref{e2m}) are
normalized to $\balpha^2=2$ and
the factor $\2 $ in (\ref{phigen}), (\ref{Amu}) is related to conventional
normalization of the Gell-Mann matrices $\Tr(\lambda_a\lambda_b) =
\2\delta_{ab}$).

The W-boson masses (\ref{Higgs}), (\ref{wbps}) are proportional to
\be
\phi_{ij}\equiv \phi_i-\phi_j = \balpha\bphi
\ee
are given then by the scalar products with the {\em roots} so that the (perturbative)
prepotential for pure gauge theory may be rewritten as
\be
\label{pp3}
{\cal F} = \2\sum_{\balpha\in\Delta_+}(\balpha\bphi)^2\log {\balpha\bphi\over\Lambda}
\ee
where sum is taken over all positive roots $\Delta_+$. The quadratic part of (\ref{pp3})
can be easily rewritten in terms of Cartesian co-ordinates or weights of the
fundamental representation, using
\be
\bphi^2 = {1\over C_V}\sum_{\balpha\in\Delta_+}(\balpha\bphi)^2 =
\sum_{\bmu}\left(\bmu\bphi\right)^2
\ee
or
\be
\bphi^2 = {1\over N_c}\sum_{i<j} (\phi_{ij})^2 =
{1\over 2N_c}\sum_{i,j} (\phi_{ij})^2
\stackreb{\sum_{i=1}^{N_c}\phi_i=0}{=} \sum_{i=1}^{N_c}\phi_i^2 =
\2\sum_{i=1}^{N_c-1} a_i^2
\ee
The matrix of coupling constants for the $SU(3)$ gauge theory may be easily computed
in terms of the root variables
${\cal A}_1 = \phi_1-\phi_3 = \balpha_{12}\bphi$ and
${\cal A}_2 = \phi_2-\phi_3 = \balpha_2\bphi$,
where we used one of the simple roots $\balpha_2$ and the "highest" root
$\balpha_{12} = \balpha_1+\balpha_2$ (see fig.~\ref{fi:su3})
\be
\label{troot3}
T_{ij} = {\d{\cal F}\over\d {\cal A}_i\d {\cal A}_j} =
\left(
\begin{array}{cc}
\log{{\cal A}_1\over\Lambda} + \log{{\cal A}_{12}\over\Lambda} + 3 &
- \log{{\cal A}_{12}\over\Lambda} \\
- \log{{\cal A}_{12}\over\Lambda} &
\log{{\cal A}_2\over\Lambda} + \log{{\cal A}_{12}\over\Lambda} + 3
\end{array}\right)
\ee
and it can be rewritten in following co-ordinates
\be
\label{qed2}
{\d{\cal F}\over\d a_i\d a_j} =
\left(
\begin{array}{cc}
T_{11}+T_{22}-2T_{12} &
\sqrt{3}\left(T_{11}-T_{22}\right) \\
\sqrt{3}\left(T_{11}-T_{22}\right) &
3\left(T_{11}+T_{22}\right)+6T_{12}
\end{array}\right) =
\\
= \left(
\begin{array}{cc}
\log{{\cal A}_1{\cal A}_2{\cal A}_{12}^4\over\Lambda^6} + 6 &
\sqrt{3}\log{{\cal A}_1\over {\cal A}_2} \\
\sqrt{3}\log{{\cal A}_1\over {\cal A}_2} &
3\log{{\cal A}_1{\cal A}_2\over\Lambda^3} + 18
\end{array}\right)
\ee
If all ${\cal A}_i\sim M \to\infty$
\be
{\d{\cal F}\over\d a_i\d a_j} \sim 6\log M
\left(
\begin{array}{cc}
1 & 0 \\
0 & 1
\end{array}\right) + {\cal O}(1)
\ee
with the coefficient $\beta_{YM}=2N_c=6$.

One may also consider the contribution of the fundamental multiplets to the set
of effective constants (\ref{troot3}). They come from the contribution of
fundamental massive multiplets to the prepotential
\be
{\cal F}_f = - {1\over 4}\sum_{A=1}^{N_f}\sum_{\bmu}({\bmu\bphi + m_A})^2
\log {\bmu\bphi + m_A\over\Lambda}
\ee
whose second derivatives w.r.t. $a_i$ give rise to the contribution to
(\ref{troot3}) of the form
\be
{\d^2{\cal F}_f\over\d {\cal A}_i\d {\cal A}_j} = {1\over 18}\sum_{A=1}^{N_f}
\left(
\begin{array}{cc}
-\log {(B_1^A)^4B_2^AB^A\over\Lambda^6} &  \log {(B_1^A)^2(B_2^A)^2\over B^A\Lambda^3}\\
\log {(B_1^A)^2(B_2^A)^2\over B^A\Lambda^3} &  -\log {B_1^A(B_2^A)^4B^A\over\Lambda^6}
\end{array}\right) + const
\ee
where
\be
B_1^A = \bmu_1\bphi + m_A = {2\over 3}{\cal A}_1 - {1\over 3}{\cal A}_2 + m_A
\\
B_2^A = \bmu_2\bphi + m_A = {2\over 3}{\cal A}_2 - {1\over 3}{\cal A}_1 + m_A
\\
B^A = - \bmu_3\bphi - m_A = {1\over 3}{\cal A}_1 + {1\over 3}{\cal A}_2 - m_A
\ee
In following co-ordinates one gets
\be
\label{t3fun}
{\d^2{\cal F}_f\over\d a_i\d a_j} = {1\over 18}\sum_{A=1}^{N_f}
\left(
\begin{array}{cc}
-\log{(B_1^A)^9(B_2^A)^9\over\Lambda^{18}} &  \sqrt{3}\log {(B_2^A)^3\over
(B_1^A)^3}\\
\sqrt{3}\log {(B_2^A)^3\over (B_1^A)^3} &  -3
\log {B_1^AB_2^A(B^A)^4\over\Lambda^6}
\end{array}\right) + const
\ee
In the limit ${\cal A}_i \sim M \to\infty$ (\ref{t3fun}) gives rise to
\be
{\d^2{\cal F}_f\over\d a_i\d a_j} = - N_f\log M
\left(
\begin{array}{cc}
1 & 0 \\
0 & 1
\end{array}\right) + {\cal O}(1)
\ee
with the coefficient giving necessary contribution into $\beta_{QCD} =
2N_c-N_f$.

\end{document}